\documentclass[cernpreprint, atlasdraft=false, USenglish, texmf, orcidlogo]{atlasdoc}
\usepackage{atlaspackage}
\usepackage{multirow}
\usepackage[flushleft]{threeparttable}
\usepackage{adjustbox}
\usepackage[bottom]{footmisc}
\usepackage{atlasbiblatex}

\usepackage[jetetmiss=true, particle=true, process=true, bsm=true]{atlasphysics}
\graphicspath{{logos/}{figures/}}

\usepackage{ANA-HMBS-2024-11-PAPER-defs}

\AtlasTitle{Search for cascade decays of charged sleptons and sneutrinos in final states with three leptons and missing transverse momentum in $pp$ collisions at $\sqrt{s}=13$ TeV with the ATLAS detector}

\AtlasAbstract{%
A search for cascade decays of charged sleptons and sneutrinos using final states characterized by three leptons (electrons or muons) and missing transverse momentum is presented.
The analysis is based on a dataset with 140~\(\text{fb}^{-1}\) of proton--proton ($pp$) collisions at a center-of-mass energy of \(\sqrt{s}=13\)~TeV recorded by the ATLAS detector at the Large Hadron Collider.
This paper focuses on a supersymmetric scenario that is motivated by the muon anomalous magnetic moment observation, dark mattter relic density abundance, and electroweak naturalness.
A mass spectrum involving light higgsinos and heavier sleptons with a bino at intermediate mass is targeted.
No significant deviation from the Standard Model expectation is observed.
This search enables to place stringent constraints on this model, excluding at the 95\% confidence level charged slepton and sneutrino masses up to 450~GeV when assuming a lightest neutralino mass of 100~GeV and mass-degenerate selectrons, smuons and sneutrinos.
}

\AtlasRefCode{HMBS-2024-11}

\PreprintIdNumber{CERN-EP-2025-045}

\AtlasJournalRef{Phys. Rev. D 112 (2025) 012005}
\AtlasDOI{10.1103/6gy3-cb4t}

\AtlasCoverEgroupAnalysisTeam{atlas-hmbs-2024-11-analysis-team@cern.ch}

\hypersetup{pdftitle={ATLAS document},pdfauthor={The ATLAS Collaboration}}

\begin{document}

\maketitle

\tableofcontents

\section{Introduction}
%

Supersymmetry (SUSY)~\cite{Golfand:1971iw,Volkov:1973ix,Wess:1974tw,Wess:1974jb,Ferrara:1974pu,Salam:1974ig} postulates a symmetry between bosons and fermions, and predicts the existence of new partners (`superpartners') for each Standard Model (SM) particle.
In SUSY models conserving \(R\)-parity~\cite{Farrar:1978xj}, SUSY particles are produced in pairs.
The lightest supersymmetric particle (LSP) has to be stable and is weakly interacting, constituting a possible dark-matter candidate~\cite{Goldberg:1983nd,Ellis:1983ew}.
The LSP produced at the Large Hadron Collider (LHC)~\cite{Evans:2008zzb} would escape detection and cause momentum imbalance in the form of missing transverse momentum (\Ptmiss, the magnitude of which is referred to as \met) in the final state, which is used to discriminate the SUSY signal from the background.

The scalar superpartners of the SM fermions are charged sleptons (\slepton), sneutrinos (\snu), and squarks (\squark), while gluons have fermionic superpartners called gluinos (\gluino).
In the minimal supersymmetric extension of the SM (MSSM)~\cite{Fayet:1976et,Fayet:1977yc}, the bino, wino and higgsino fields are the fermionic superpartners of the \(\mathrm{SU(2)}\times\mathrm{U(1)}\) gauge fields in the SM, and the two complex scalar doublets of a minimally extended Higgs sector, respectively.
The bino, wino, and higgsino are collectively referred to as `electroweakinos', and they mix to give the mass eigenstates referred to as charginos \(\widetilde{\chi}^{\pm}_i\,(i=1,2)\) and neutralinos \(\widetilde{\chi}^0_j\,(j=1,2,3,4)\), with the subscripts indicating increasing mass.

\begin{figure}[t!]
\centering
\includegraphics[width=0.30\textwidth]{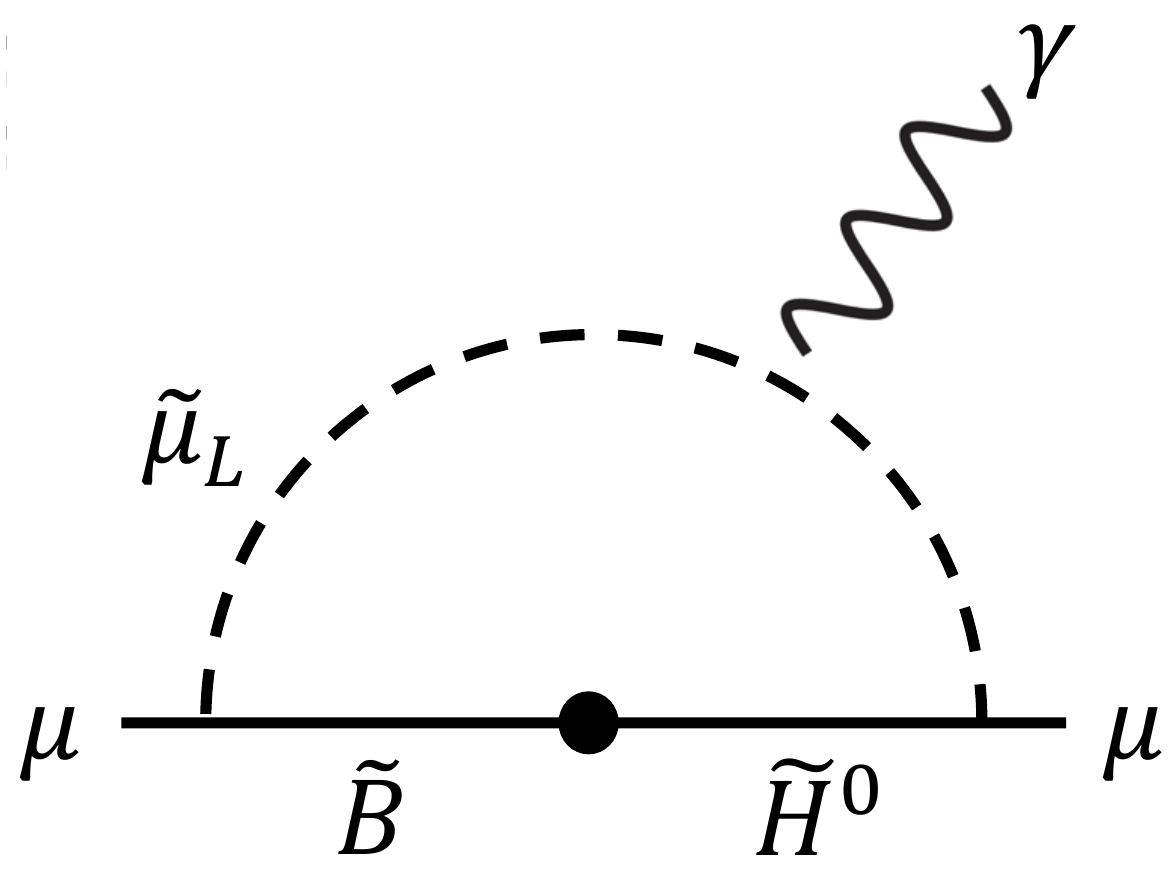}
\caption{The neutralino-smuon loop yielding a sizable muon anomalous magnetic moment.
The black dot represents the mixing between electroweakinos.
}
\label{fig:loop}
\end{figure}

Electroweakinos and sleptons with masses of several hundred \GeV are motivated by various phenomenological arguments: the MSSM parameter space explaining the possible discrepancy between the measured muon anomalous magnetic moment~\cite{Aguillard:2023} and its SM predictions~\cite{Aoyama:2020ynm}
\footnote{Note that the discrepancy tends to smaller when the lattice QCD results are used for the thoery prediction~\cite{colangelo2022prospects}.}
typically includes electroweakinos and smuons with masses from 200~\GeV to 1~\TeV~\cite{moroi1996muon,PhysRevLett.86.3484,endo2020muon}; when the neutralino LSP is the dark matter candidate, its mass is constrained to be less than a few \TeV by the observed relic density~\cite{HISANO200734,PhysRevD.98.035026}; the higgsino mass is also motivated to be of the same order as the \Zboson-boson mass by electroweak naturalness arguments~\cite{Barbieri:2009ev,Baer:2011ec,Papucci:2011wy,Baer:2012up}.
The additional muon anomalous magnetic moment contribution is typically generated through chargino-sneutrino loops and neutralino-smuon
loops containing at least three types of supersymmetric particles~\cite{moroi1996muon,endo2020muon}.

This paper targets the loop contribution including the bino, higgsino, and left-handed smuon as shown in Figure~\ref{fig:loop},
and specifically the mass spectrum involving the higgsino LSP, heavier left-handed sleptons and sneutrinos, and the bino at a mass in between, as illustrated in Figure~\ref{fig:decaypattern}.
Higgsino-dominated states (\Ntwo, \Cone, \None) have a mass-compressed spectrum in this model.
This mass spectrum is referred to as the `slepton-bino-higgsino' model (`SBH' model) in this paper,
and can address the observed muon anomalous magnetic moment, dark matter, and electroweak naturalness simultaneously when the involved SUSY particles are lighter than 1~\TeV~\cite{moroi1996muon}.
While the SBH model involving smuons is more motivated by the muon anomalous magnetic moment measurement, a model with selectrons is also considered in this analysis as the electron anomalous magnetic moment measurements~\cite{PhysRevLett.130.071801,PhysRevLett.100.120801} may also imply a mild deviation from the SM prediction~\cite{atoms7010028,PhysRevD.97.036001,cite-key,Parker_2018}.
Current constraints on the SBH model primarily originate from direct dark matter searches using nuclear recoils~\cite{PhysRevLett.131.041002} and SUSY searches at the LHC.
The direct dark matter searches place a stringent constraint for \(\Delta m(\Nthree,\None)\lesssim100~\GeV\), while \(\Delta m(\Nthree,\None)\gtrsim100~\GeV\) is still fully viable assuming a local dark matter density of $0.3~\GeV \cdot \mathrm{cm}^{-3}$~\cite{EPJC.81.907.2021}.

At the LHC, no dedicated searches have been performed for the SBH model, however some constraints can be set through
(i) generic higgsino LSP searches, using the disappearing track signature~\cite{SUSY-2018-19,CMS-EXO-16-044,CMS-EXO-19-010},
the mildly displaced track signature~\cite{SUSY-2020-04},
and the low-momentum prompt lepton\footnote{In this paper, `leptons' refer to electrons or muons.} signatures~\cite{SUSY-2016-25,SUSY-2019-09,CMS-SUS-19-012};
(ii) several electroweakino and slepton searches using final states with two or three leptons
and \met~\cite{SUSY-2018-32,SUSY-2013-11,SUSY-2016-24,SUSY-2019-02,CMS-SUS-13-006,CMS-SUS-17-009, SUSY-2018-36,SUSY-2018-32,SUSY-2019-09}.
However, their sensitivity to the SBH model is mostly limited to the mass parameter space that is already disfavored by the direct dark matter searches, i.e. \(\Delta m(\Nthree,\None)\lesssim 100~\GeV\).
In the SBH model, sleptons rarely decay into higgsinos directly due to the suppressed slepton-higgsino Yukawa coupling compared with the slepton-bino electroweak coupling.

This study reports the first dedicated search for the SBH model targeting the cascade decay signature using the proton--proton collision data at a center-of-mass energy of $\sqrt{s}=13~\TeV$ collected by the ATLAS detector at the LHC in the years 2015--2018, corresponding to an integrated luminosity of 140~\ifb~\cite{DAPR-2021-01}.
Final states with three leptons are explored.
Through the optimization of the event selection, and the introduction of new search regions requiring three leptons with the same charge and zero jets containing \(b\)-hadrons, a signature explored in this paper for the first time at the LHC, the first unique sensitivity for \(\Delta m(\Nthree,\None)\gtrsim 100~\GeV\) is achieved.

The paper is structured as follows;
the definition of the benchmark signal model is further detailed in Section~\ref{sec:model};
a brief overview of the ATLAS detector is provided in Section~\ref{sec:ATLAS};
the data and the Monte Carlo (MC) simulation samples are described in Section~\ref{sec:samples};
the particle reconstruction methods used in the analysis are presented in Section~\ref{sec:object};
the event selection strategy and the signal region definition are discussed in Section~\ref{sec:selection};
the SM background estimation is described in Section~\ref{sec:background}, followed by a summary of the systematic uncertainties in Section~\ref{sec:systematics};
the results of the search and its interpretation are presented in Section~\ref{sec:result},
followed by the conclusion in Section~\ref{sec:conclusion}.

\begin{figure}[t!]
\centering
\includegraphics[width=0.3\textwidth]{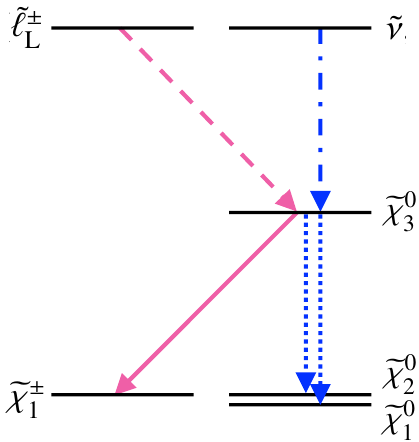}
\caption{The mass spectrum and the decay pattern of the considered simplified SBH model.
The dashed (dashdotted) arrow represents the decay emitting a charged lepton (neutrino).
The solid (dotted) arrow represents the decay emitting a \Wboson (\Zboson or \(h\)) boson.
Light pink (darker blue) arrows represent the decay emitting a charged (neutral) particle.
}
\label{fig:decaypattern}
\end{figure}

\section{The SBH signal model} \label{sec:model}
The benchmark signal model targeted in the analysis involves the direct pair production of left-handed sleptons and sneutrinos (\(\sleptonL^+\sleptonL^-\), \(\snu\snu\)),
or the associated production of a left-handed slepton and a sneutrino (\(\sleptonL^{\pm}\snu\)) mediated by off-shell \Wboson/\Zboson bosons.

Sleptons and sneutrinos promptly decay into \(\sleptonL\rightarrow\ell\Nthree\) and \(\snu\rightarrow\nu\Nthree\),
and the \Nthree further decays as \(\Nthree\rightarrow\Wboson^{\pm}\Conemp\), \(\Nthree\rightarrow\Zboson\Nonetwo\) or \(\Nthree\rightarrow h\Nonetwo\), as illustrated in Figure~\ref{fig:decaypattern}.
The decay of the \Nthree via a Higgs boson, $h$, is only kinematically allowed when the mass difference between the bino-dominated state (\Nthree) and the higgsino-dominated states (\Cone, \Ntwo) is larger than the Higgs-boson mass.
Example signal diagrams with three leptons in the final state are illustrated in Figure~\ref{fig:Diagram}.
Right-handed sleptons are not considered in this model due to their much smaller production cross-section.
However, they can be targeted by the same search since they follow the same decay chains as the left-handed ones.
The presence of right-handed sleptons would also not influence the decay chains of the left-handed sleptons and sneutrinos.

Leptons can originate from the decay of sleptons, the direct decay of \Wboson/\Zboson bosons, or the indirect decay of a Higgs boson (mostly via $h \rightarrow WW^*$).
The analysis focuses on final states with three leptons as they are found to be the most sensitive compared with the other lepton multiplicity categories.
The signatures are also characterized by the presence of \met originating from the LSPs and neutrinos.

\begin{figure}[t!]
\centering
\subfloat[]{
\includegraphics[height=0.27\textwidth, keepaspectratio]{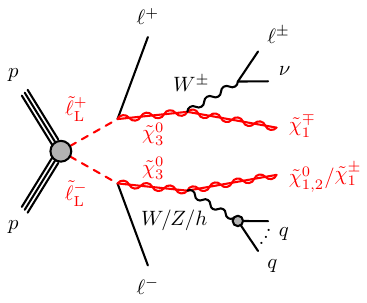}
\label{fig:slslWV}
}
\subfloat[]{
\includegraphics[height=0.27\textwidth, keepaspectratio]{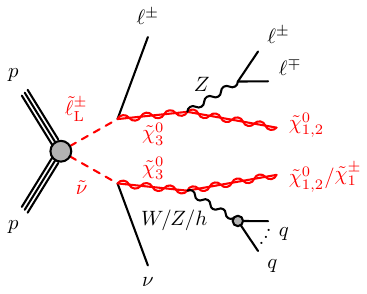}
\label{fig:slsnZV}
}
\subfloat[]{
\includegraphics[height=0.27\textwidth, keepaspectratio]{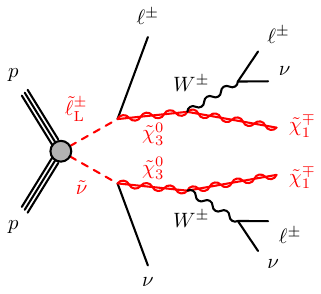}
\label{fig:slsnWW}
}
\caption{Diagrams for the targeted cascade signatures in the SBH model. Final states with three light-flavor charged leptons (electrons or muons) are considered in the analysis. The decays of \Ntwo and \Cone are ignored as they only result in low-momentum particles that are almost never reconstructed.\label{fig:Diagram}}
\end{figure}

The analysis design and the result interpretation are based on the framework of `simplified models'~\cite{Alwall:2008ag},
where the masses of relevant SUSY particles (in this case the \sleptonL, \snu, \Nthree, \Ntwo, \Cone, and \None) are the only free parameters.
For the sleptons and sneutrinos, three sub-scenarios are considered:
\begin{itemize}
\item $m(\selL)=m(\snu_e) < 1~\TeV$ with $\smuL, \snu_\mu$ being decoupled,
\item $m(\smuL)=m(\snu_\mu) < 1~\TeV$ with $\selL, \snu_e$ being decoupled,
\item $m(\selL)=m(\snu_e)=m(\smuL)=m(\snu_\mu) < 1~\TeV$.
\end{itemize}
Staus and right-handed sleptons are not considered.
The mass splitting between the left-handed sleptons and sneutrinos, and between the higgsino-dominated states (\Ntwo, \Cone, \None) with non-negligible bino mixing, is typically \(\mathcal{O}(1-10)~\GeV\)~\cite{Martin:1997ns,PhysRevD.109.095045}.
The states are treated as mass-degenerate in this study, since the low-\pT particles from decays in such compressed hierarchies are rarely reconstructed in the analysis.
A 100\% branching ratio is assumed for \(\sleptonL\rightarrow\ell\Nthree\) and \(\snu\rightarrow\nu\Nthree\), which is reasonable given the coupling assumptions.
The branching ratios of \Nthree are set to:
\begin{itemize}
\item \(\mathcal{B}(\Nthree\rightarrow\Wboson^{\pm(*)}\Conemp)=50\%\,\,(50\%)\),
\item \(\mathcal{B}(\Nthree\rightarrow\Zboson^{(*)}\Nonetwo) = 25\%\,\,(50\%)\),
\item \(\mathcal{B}(\Nthree\rightarrow h\Nonetwo) = 25\%\,\,(0\%)\),
\end{itemize}
with equal branching ratios into \None and \Ntwo, for \(\Delta m(\Nthree, \None)\geq m_h\) (\(\Delta m(\Nthree, \None)< m_h\)).
The values for \(\Delta m(\Nthree, \None)\geq m_h\) are typical for these masses~\cite{PhysRevD.109.015027}.
The values for \(\Delta m(\Nthree, \None)< m_h\) are set more arbitrarily, but this choice has a negligible impact on the search results.
Other SUSY particles are decoupled.


%

%
\section{The ATLAS detector}
\label{sec:ATLAS}

The ATLAS detector~\cite{PERF-2007-01} at the LHC covers nearly the entire solid angle around the collision point.\footnote{ATLAS uses a right-handed coordinate system with its origin at the nominal interaction point (IP) in the center of the detector and the \(z\)-axis along the beam pipe. The \(x\)-axis points from the IP to the center of the LHC ring, and the \(y\)-axis points upwards. Polar coordinates \((r, \phi)\) are used in the transverse plane, \(\phi\) being the azimuthal angle around the \(z\)-axis. The pseudorapidity is defined in terms of the polar angle \(\theta\) as \(\eta=-\ln\tan(\theta/2)\) and is equal to the rapidity \(y=\frac{1}{2}\ln\left(\frac{E+p_z}{E-p_z}\right)\) in the relativistic limit. Angular distance is measured in units of \(\Delta R\equiv\sqrt{(\Delta y)^2+(\Delta\phi)^2}\).}
It consists of an inner tracking detector surrounded by a thin superconducting solenoid, electromagnetic and hadron calorimeters, and a muon spectrometer incorporating three large superconducting air-core toroidal magnets.

The inner-detector system (ID) is immersed in a \(2~\mathrm{T}\) axial magnetic field and provides charged-particle tracking in the range of \(\abseta<2.5\).
The high-granularity silicon pixel detector covers the vertex region and typically provides four measurements per track, the first hit normally being in the insertable B-layer (IBL) installed before Run~2~\cite{ATLAS-TDR-19,PIX-2018-001}.
It is followed by the SemiConductor Tracker (SCT), which usually provides eight measurements per track.
These silicon detectors are complemented by the transition radiation tracker (TRT), which enables radially extended track reconstruction up to \(\abseta<2.0\).
The TRT also provides electron identification information based on the fraction of hits (typically 30 in total) above a higher energy-deposit threshold corresponding to transition radiation.

The calorimeter system covers the pseudorapidity range \(\abseta<4.9\).
Within the region \(\abseta<3.2\), electromagnetic calorimetry is provided by barrel and endcap high-granularity lead/liquid-argon (LAr) calorimeters, with an additional thin LAr presampler covering \(\abseta<1.8\) to correct for energy loss in material upstream of the calorimeters.
Hadron calorimetry is provided by the steel/scintillator-tile calorimeter, segmented into three barrel structures within \(\abseta<1.7\), and two copper/LAr hadron endcap calorimeters.
The solid angle coverage is completed with forward copper/LAr and tungsten/LAr calorimeter modules optimized for electromagnetic and hadronic energy measurements respectively.

The muon spectrometer (MS) comprises separate trigger and high-precision tracking chambers measuring the deflection of muons in a magnetic field generated by the superconducting air-core toroidal magnets.
The field integral of the toroids ranges between 2.0 and \(6.0~\mathrm{T\,m}\) across most of the detector.
Three layers of precision chambers, each consisting of layers of monitored drift tubes, cover the region \(\abseta<2.7\), complemented by cathode-strip chambers in the forward region, where the background is higher.
The muon trigger system covers the range \(\abseta<2.4\) with resistive-plate chambers in the barrel, and thin-gap chambers in the endcap regions.

The luminosity is measured mainly by the LUminosity Cherenkov Integrating Detector 2 (LUCID-2)~\cite{LUCID2} detector that records Cherenkov light produced by the quartz windows of photomultipliers located close to the beam pipe.

Events are selected by the first-level trigger system implemented in custom hardware, followed by selections made by algorithms implemented in software in the high-level trigger~\cite{TRIG-2016-01}.
The first-level trigger accepts events from the \(40~\unit{\MHz}\) bunch crossings at a rate below \(100~\unit{\kHz}\), which the high-level trigger further reduces to record events to disk at about \(1~\unit{\kHz}\).

A software suite~\cite{SOFT-2022-02} is used in data simulation, in the reconstruction and analysis of real and simulated data, in detector operations, and in the trigger and data acquisition systems of the experiment.


%

%
\section{Data and Monte Carlo simulation}
\label{sec:samples}

The analysis is performed using the \pp collision data collected by the ATLAS detector at the LHC between the years 2015 and 2018 at a center-of-mass energy of \(\sqrt{s}=13~\TeV\).
The dataset corresponds to a total integrated luminosity of $140~\ifb$ after imposing data quality requirements~\cite{DAPR-2021-01}.
In this dataset there are, on average, approximately 34 simultaneous $pp$ collisions in each bunch crossing.

Monte Carlo simulation is used to model the contributions of the signal and the SM processes.
It is used to define and optimize the event selection criteria, to estimate the signal and SM background event yields after the selections, and to evaluate the systematic uncertainties associated with the estimation.
The generators and parameters used in the MC simulation samples are described below and summarized in Table~\ref{tab:MCSummary}.

\begin{table}[t!]
\centering
\caption{Summary of the generator configurations for the simulated SM backgrounds and signal samples.
}
{\tiny
\begin{tabular*}{\textwidth}{@{\extracolsep{\fill}}c|ccccc}
\toprule
Process & Matrix element & Parton shower & Tune & PDF set & Cross-section order \\ \midrule\midrule
SUSY processes & \MGNLO[3.3.1]~\cite{Alwall:2014hca} & \PYTHIA[8.307]~\cite{Bierlich:2022pfr} & A14~\cite{ATL-PHYS-PUB-2014-021,ATL-PHYS-PUB-2017-007} & \NNPDF[2.3lo]~\cite{Ball:2012cx} & NLO+NLL~\cite{Debove:2010kf,Fuks:2012qx,Fuks:2013vua,Fiaschi:2018hgm,Bozzi:2007qr,Fuks:2013lya,Fiaschi:2018xdm} \\
\(W/Z\text{\,+\,jets}\) ~\cite{ATL-PHYS-PUB-2017-006} & \SHERPA 2.2.11/2.2.14~\cite{Bothmann:2019yzt} & \SHERPA 2.2.11/2.2.14 & standard & \NNPDF[3.0nnlo]~\cite{Ball:2014uwa} & NNLO~\cite{Anastasiou:2003ds} \\
Diboson~\cite{ATL-PHYS-PUB-2017-005} & \SHERPA 2.2.11/2.2.12 & \SHERPA 2.2.11/2.2.12 & standard & \NNPDF[3.0nnlo] & Generator NLO \\
Triboson~\cite{ATL-PHYS-PUB-2017-005} & \SHERPA 2.2.2 & \SHERPA 2.2.2 & standard & \NNPDF[3.0nnlo] & Generator NLO \\
\ttbar~\cite{Frixione:2007nw} & \POWHEGBOX 2~\cite{Nason:2004rx,Frixione:2007vw,Alioli:2010xd} & \PYTHIA[8.230] & A14 & \NNPDF[2.3lo] & NNLO+NNLL~\cite{Beneke:2011mq,Cacciari:2011hy,Baernreuther:2012ws,Czakon:2012zr,Czakon:2012pz,Czakon:2013goa,Czakon:2011xx}  \\
Single-top~\cite{Re:2010bp,Frederix:2012dh,Aliev:2010zk,Kant:2014oha} & \POWHEGBOX 2 & \PYTHIA[8.230]~\cite{Sjostrand:2014zea} & A14 & \NNPDF[2.3lo] & NNLO+NNLL~\cite{Kidonakis:2010ux,Kidonakis:2013zqa,Aliev:2010zk,Kant:2014oha,}\\
\(\ttbar W\) & \SHERPA 2.2.10 & \SHERPA 2.2.10 & standard & \NNPDF[3.0nnlo] & Generator NLO \\
\(\ttbar Z, tZ, tWZ\) & \MGNLO[2.3] & \PYTHIA[8.210] & A14 & \NNPDF[3.0nnlo] & Generator NLO \\
\(\ttbar WW\), 3-top, 4-top & \MGNLO[2.2] & \PYTHIA[8.186]~\cite{Sjostrand:2007gs} & A14 & \NNPDF[2.3lo] & Generator LO \\
\(\ttbar h\)~\cite{Hartanto:2015uka} & \POWHEGBOX 2 & \PYTHIA[8.230] & A14 & \NNPDF[2.3lo] & Generator NLO \\
Higgs (ggF) & \POWHEGBOX 2 & \PYTHIA[8.212] & AZNLO~\cite{STDM-2012-23} & CTEQ6L1~\cite{Pumplin:2002vw} & NNNLO+NLO(EWK)~\cite{deFlorian:2016spz,Anastasiou:2016cez,Anastasiou:2015ema,Dulat:2018rbf,Aglietti:2004nj,Actis:2008ug,Bonetti:2018ukf} \\
Higgs (\(Vh\)) & \POWHEGBOX 2 & \PYTHIA[8.230] & AZNLO & CTEQ6L1 & NNLO+NLO(EWK)~\cite{deFlorian:2016spz} \\
\bottomrule
\end{tabular*}
}
\label{tab:MCSummary}
\end{table}

Events from \(W/Z\text{\,+\,jets}\), diboson and triboson processes~\cite{ATL-PHYS-PUB-2017-005,ATL-PHYS-PUB-2017-006} are simulated using the \SHERPA[2.2]\cite{Bothmann:2019yzt} generator:
the fully leptonically decaying diboson events are simulated using \SHERPA[2.2.12];
events of \Zee+\,jets, \Zmm+\,jets, \Wjets, and semileptonically decaying diboson processes are simulated with \SHERPA[2.2.11];
\Ztt samples are simulated with \SHERPA[2.2.14]; triboson processes are simulated using \SHERPA[2.2.2].
The matrix element calculations are matched to the parton shower (PS) simulation using the Catani--Seymour dipole factorization~\cite{Gleisberg:2008fv,Schumann:2007mg}.
This matching is performed separately for different jet multiplicities and merged into an inclusive sample using an improved Catani-Krauss-Kuhn-Webber (CKKW) matching procedure~\cite{Catani:2001cc,Hoeche:2009rj} extended to next-to-leading-order (NLO) accuracy in QCD, using the \MEPSatNLO prescription~\cite{Hoeche:2011fd,Hoeche:2012yf,Catani:2001cc,Hoeche:2009rj}.
The virtual QCD correction for matrix elements at NLO accuracy is provided by the \openloops library~\cite{Cascioli:2011va,Denner:2016kdg}.
Virtual electroweak loop-terms are included at NLO accuracy for the \(W/Z\text{\,+\,jets}\) and diboson processes.
The \NNPDF[3.0nnlo]~\cite{Ball:2014uwa} set of parton distribution functions (PDFs) is used together with a dedicated set of tuned PS parameters (`tune') developed by the \SHERPA authors~\cite{Schumann:2007mg}.
The \(W/Z\text{\,+\,jets}\) (diboson) samples are calculated for up to two (one) additional partons at NLO and up to four (three) additional partons at leading-order (LO) in QCD, and the triboson samples are calculated at NLO in QCD for the inclusive processes and at LO in QCD for up to two additional parton emissions.
The diboson samples include the loop-induced and electroweak production.
The Higgs boson contributions are not included in the diboson and triboson samples.
The cross-sections calculated by the event generators are used for all these samples except for \(W/Z\text{\,+\,jets}\), which are normalized to a next-to-next-to-leading-order (NNLO) cross-section prediction~\cite{Anastasiou:2003ds}.

The \ttbar, \ttH and the single-top \(t\)-channel, \(s\)-channel and \tW processes are modeled using \POWHEGBOX[2] \(+\) \PYTHIA[8.230]~\cite{Nason:2004rx,Frixione:2007vw,Alioli:2010xd,Sjostrand:2014zea}.
The \hdamp parameter\footnote{The \hdamp parameter is a resummation damping factor and one of the parameters that controls the matching of \textsc{Powheg} matrix elements to the parton shower and thus effectively regulates the high-\pT radiation against which the \ttbar system recoils.} is set to 1.5 times the top-quark mass~\cite{ATL-PHYS-PUB-2016-020}.
The samples are generated employing the five-flavor scheme (four-flavor for the single-top \(t\)-channel), and a diagram removal scheme~\cite{Frixione:2008yi} is used to remove the interference and overlap of the \tW process with the \ttbar production.
The \ttW process is simulated using the \SHERPA[2.2.10] generator.
The matrix elements are calculated for up to one additional parton at NLO and up to two partons at LO using \comix~\cite{Gleisberg:2008fv} and \openloops,
and merged with the \SHERPA parton shower using the \MEPSatNLO prescription with a merging scale of 30~\GeV.
Other top-quark-involved processes (\ttZ, \tZ, \tWZ, \ttWW, 3-top and 4-top) are modeled using \MGNLO[2] \(+\) \PYTHIA[8]~\cite{Alwall:2014hca,Sjostrand:2007gs}.
Samples of Higgs boson production via gluon--gluon fusion and associated production are simulated using \POWHEGBOX[2] \(+\) \PYTHIA[8].

The SUSY signal production is generated with LO matrix elements with up to two extra partons using \MGNLO[3.3.1] interfaced with \PYTHIA[8.307]~\cite{Bierlich:2022pfr} with the A14~\cite{ATL-PHYS-PUB-2014-021} tune to perform the SUSY particle decays, parton showering, hadronization, and the underlying event simulation.
The parton luminosities used are provided by the \NNPDF[2.3lo] PDF set~\cite{Ball:2012cx}.
Jet-parton matching is performed following the CKKW-L prescription~\cite{Lonnblad:2001iq,Lonnblad:2011xx}, with a matching scale set to one quarter of the mass of the pair-produced SUSY particles.
The generated signal events are required to have at least two leptons.
The signal cross-sections are calculated up to NLO in \alphas adding the resummation of soft gluon emission at next-to-leading-logarithm accuracy (NLO+NLL)~\cite{Debove:2010kf,Fuks:2012qx,Fuks:2013vua,Fiaschi:2018hgm,Bozzi:2007qr,Fuks:2013lya,Fiaschi:2018xdm}.
The nominal cross-sections and their uncertainties are taken from an envelope of cross-section predictions using different PDF sets and factorization and renormalization scales, as described in Ref.~\cite{Borschensky:2014cia}.
For example, for a slepton/sneutrino mass of \(400~\GeV\), the production cross-sections for \(\sleptonL^+\sleptonL^-\) and \(\sleptonL^{\pm}\snu\) are
\(1.33\pm0.04~\unit{\fb}\) and \(4.82\pm0.19~\unit{\fb}\) respectively, for each generation of the left-handed slepton.

The simulation of $b$- and $c$-hadron decays in the samples generated with \POWHEGBOX or \MGNLO is performed with \EVTGEN~1.6.0~\cite{Lange:2001uf}.

All MC events are propagated through a full simulation of the ATLAS detector~\cite{SOFT-2010-01} using \textsc{Geant4}~\cite{Agostinelli:2002hh} to model the interactions of particles with the detector, except those from the SUSY signal and 4-top processes for which a parameterized simulation of the ATLAS calorimeter~\cite{SOFT-2010-01,SIMU-2018-04} is used.
The effect of multiple interactions in the same and neighboring bunch crossings (pileup) is modeled by overlaying simulated minimum-bias collisions onto each hard-scattering event.
The minimum-bias events are generated with \PYTHIA[8.186] using the A3 tune~\cite{ATL-PHYS-PUB-2016-017} and \MSTW{LO} PDF set~\cite{Martin:2009iq}.
For each simulated hard-scatter process a separate MC sample is generated to reflect the conditions of the 2015+2016, the 2017, and the 2018 datasets.
The number of overlaid minimum-bias collisions is sampled for each event according to the distribution of the average number of interactions per bunch crossing measured in that dataset.


%

%
\section{Event reconstruction}
\label{sec:object}

%
The collision data were collected with triggers requiring at least a single electron or a single muon reconstructed by the trigger system, with various lepton transverse momentum (\pt) thresholds depending upon their relative quality including isolation~\cite{TRIG-2018-05,TRIG-2018-01,TRIG-2016-01}.
To ensure trigger efficiencies are well understood in the analysis phase space, tighter quality and \pT requirements are applied to fully reconstructed `signal' leptons, which are defined below.
Each event for which the trigger was activated is required to have at least one electron (muon) with a fully calibrated \pt above 27, 61, or 141~\GeV (27.3 or 52.5~\GeV), with larger-\pt requirements corresponding to reduced lepton-quality requirements of the trigger.
For the 2015 data, the \pt requirement is lowered to 26, 61, or 121~\GeV (21 or 52.5~\GeV) for the electron (muon).

Both the data and MC events are required to have at least one reconstructed vertex that is associated with two or more tracks of transverse momentum $\pt > 500~\MeV$.
The primary vertex of each event is selected as the vertex with the largest $\Sigma\pt^2$ of its associated tracks, and used as the pivot of object reconstructions~\cite{ATL-PHYS-PUB-2015-026}.

The primary objects considered in this analysis are electrons, muons, and jets.
Two levels of selection criteria are defined for leptons and jets; the looser `baseline' criteria and the tighter `signal' criteria.
Baseline objects are used for resolving ambiguities between overlapping objects and calculating \Ptmiss.
Baseline objects are also used as inputs to the data-driven estimation of \fake leptons, which, in this paper, collectively refer to hadrons misidentified as leptons and non-prompt leptons originating from photon conversion or $b$-/$c$-hadron decays.
Tighter signal criteria are applied to the final leptons and jets considered in the analysis to ensure a high selection purity.

Baseline electrons are reconstructed from three-dimensional energy clusters in the electromagnetic calorimeter that are matched to an ID track and calibrated \insitu using $Z\rightarrow ee$ decays~\cite{EGAM-2018-01}.
In addition, baseline electrons are required to meet the `loose and B-layer likelihood' quality criteria~\cite{EGAM-2018-01}, satisfy $\pt > 4.5~\GeV$, and be within the ID acceptance ($|\eta| < 2.47$) excluding the barrel/endcap transition region of the electromagnetic calorimeter ($1.37 < |\eta| < 1.52$).

Baseline muons are reconstructed from a combined fit of tracks formed in the MS and ID, calibrated \insitu using $Z \rightarrow \mu\mu$ and $J/\psi \rightarrow \mu\mu$ decays~\cite{PERF-2015-10}, and are required to meet the `medium' quality criteria~\cite{PERF-2015-10}, satisfy $\pt > 3~\GeV$, and $|\eta|<2.5$.

Each baseline electron or muon is also required to have a trajectory consistent with the primary vertex to suppress pileup.
For this purpose, the transverse impact parameter ($d_0$) of a lepton is defined as the distance in the transverse plane between the beam-line and the closest point of the associated ID track.
The longitudinal impact parameter ($z_0$) then corresponds to the $z$-coordinate distance between that point and the primary vertex.
A selection of $|z_0 \sin\theta|<0.5~\text{mm}$, where $\theta$ is the polar angle of the track, is required for each lepton to ensure it is compatible with the primary vertex.

Baseline jets are reconstructed from particle-flow objects using the \antikt algorithm~\cite{Fastjet,Cacciari:2008gp} with a radius parameter of $R = 0.4$.
The particle-flow algorithm combines information about ID tracks and energy deposits in the calorimeters to form the input for jet reconstruction~\cite{PERF-2015-09}.
The jet energy scale (JES) and resolution (JER) are first corrected to particle level using MC simulation and then calibrated \insitu through \Zjets, \gjet, and multijet measurements~\cite{PERF-2016-04}.
Baseline jets are required to satisfy $\pt > 20$~\GeV and $|\eta| < 4.5$.
To suppress jets originating from pileup, jets with $\pT < 60$~\GeV and $|\eta| < 2.5$ are required to satisfy the
`FixedEffPt' working point of the track-based jet vertex tagger~\cite{PERF-2014-03,ATLAS-CONF-2014-018}.
The identification of baseline jets containing $b$-hadrons ($b$-jets) is performed using the `DL1dv01' multivariate discriminant built using information from track impact parameters, the presence of displaced secondary vertices, and the reconstructed flight paths of $b$- and $c$-hadrons inside the jet~\cite{FTAG-2018-01}.
The identification criteria are tuned to an average identification efficiency of 85\% as obtained for $b$-jets in simulated \ttbar events,
corresponding to rejection factors of 29, 2.6, and 3.9 for jets originating from light quarks and gluons, $c$-quarks, and $\tau$-leptons, respectively~\cite{FTAG-2018-01}.
Hadronically decaying taus are reconstructed and treated as jets in this analysis.

While photons are not used directly in the analysis, baseline photons are defined for the calculation of \Ptmiss.
Baseline photons are required to meet the `tight' quality criteria~\cite{EGAM-2018-01}, satisfy $\pt > 25~\GeV$,
and fall within the ID acceptance ($|\eta| < 2.37$) and excluding the calorimeter's transition region ($1.37 < |\eta| < 1.52$).

To prevent the reconstruction of a single particle as multiple objects, the following overlap-removal procedure is applied.
First, any electron that shares an ID track with an electron with higher \pT is removed.
Next, any electron that shares a track with a muon in the ID is removed.
Jets are removed if they are within $\Delta R = 0.2$ of an electron and are not $b$-tagged.
The remaining electrons are removed if they are within $\Delta R = \mathrm{min}(0.4, 0.04+10~\GeV/\pT(e))$ of a jet to reject fake leptons.
For the overlap of a jet with a nearby muon, the jet is discarded only if they are within $\Delta R = 0.2$ of a muon and are not $b$-tagged.
Finally, muons within $\Delta R = \mathrm{min}(0.4, 0.04+10~\GeV/\pT(\mu))$ of any remaining jets are discarded.

The \Ptmiss of each event is defined as the negative vector sum of the transverse momenta of all identified baseline objects (electrons, muons, jets, and photons), and an additional soft term constructed from all tracks associated with the primary vertex that are not associated with any baseline objects~\cite{JETM-2020-03}.
The \Ptmiss is therefore adjusted to include the full calibration of the reconstructed baseline objects while minimizing any pileup dependence in the soft term.

Signal electrons must meet the `tight' quality criteria~\cite{EGAM-2018-01}.
The track associated with each signal electron or muon must satisfy a requirement on $d_0$ and its uncertainty $\sigma_{d_0}$ such that $|d_0/\sigma_{d_0}| < 5~(3)$ for electrons (muons), ensuring the selection of leptons with prompt, well-reconstructed tracks.
In order to reduce the rate of electrons with wrongly reconstructed charge (`\chargeflip'), the \texttt{ECIDS} discriminant~\cite{EGAM-2018-01} is used, which exploits further information related to the electron track reconstruction and its compatibility with the primary vertex and the electron cluster.
Finally, signal leptons must be sufficiently well isolated from additional detector activity by satisfying a `tight' requirement on both calorimeter-based and track-based isolation variables~\cite{EGAM-2018-01,PERF-2015-10}.
Any event containing a baseline lepton that fails to meet the signal criteria is rejected to reduce the contamination from fake-lepton events.
At least one of the signal leptons must be identified as having activated a trigger and must satisfy the larger \pt requirement of that trigger.

Signal jets are required to have $|\eta| < 2.8$, and events are rejected if they contain a jet that fails to meet the `loose' quality criteria~\cite{ATLAS-CONF-2015-029}, reducing contamination from electronic noise bursts and non-collision backgrounds.
All MC simulation samples are corrected using per-event weights to account for small differences with respect to the data, in signal-lepton identification, reconstruction, isolation and triggering efficiencies~\cite{EGAM-2018-01,PERF-2015-10}, as well as in signal-jet pileup rejection~\cite{PERF-2014-03} and flavor-identification efficiencies~\cite{FTAG-2018-01}.


%

%
\section{Event selection}
\label{sec:selection}

Events with exactly three leptons are selected.
A common preselection is applied for all the signal regions (SRs) requiring the leading (sub-leading) lepton in the event to satisfy \(\pT>28\,(20)~\GeV\), while the third lepton must have \(\pT>10~\GeV\).
In addition, events with at least one \bjet are rejected in SRs to suppress the contribution of \ttbar and single-top processes.
The choice of three-lepton final states is based on the highest siginificance to the SBH model compared with two-lepton and four-lepton final states, which either have typically 200 times higher background yields or five times lower signal yields.
The lepton-\pT requirements are driven by the \pT threshold of the single-lepton trigger and the suppression of the fake-lepton backgrounds which tend to have low-\pT leptons.

Three orthogonal SRs are developed to target different signal production and decay modes, varying the requirement on the charge and flavor combination of the leptons.
A summary of the selection criteria is presented in Table~\ref{tab:SRdefinition}.

The signal region \SRSFOSon targets the \(\sleptonL^{\pm}\snu\) production followed by a \Nthree decay emitting an on-shell \Zboson boson.
At least one same-flavor opposite-charge (SFOS) lepton pair with an invariant mass \mll consistent with the \Zboson-boson mass (\(|\mll-m_Z|\leq10~\GeV\)) is required.

The signal region \SRSFOSoff targets the \(\sleptonL^+\sleptonL^-\) production followed by a \Nthree decay emitting a \Wboson boson,
and the \(\sleptonL^{\pm}\snu\) production followed by a \Nthree decay emitting an off-shell \Zboson boson resulting from a compressed mass splitting \(\Delta m(\Nthree,\None)< m_Z\).
This SR is defined by the presence of one or more SFOS lepton pair in the event with none of them having an invariant mass consistent with the \Zboson-boson mass.

The signal region \SRSS requires all three leptons to have the same charge, targeting the \(\sleptonL^{\pm}\snu\) production with the \Nthree decays resulting in a pair of same-charge \Wboson bosons.
The diagrams of the signals targeted by each SR are summarized in Figure~\ref{fig:Diagram}:
Figure~\ref{fig:Diagram}\subref{fig:slslWV} is for \SRSFOSoff, Figure~\ref{fig:Diagram}\subref{fig:slsnZV} for \SRSFOSon and Figure~\ref{fig:Diagram}\subref{fig:slsnWW} for \SRSS.

The \SRSFOSon and \SRSFOSoff (collectively referred to as \SRSFOS) are subject to the following additional selections.
A significantly large missing transverse energy, \(\met>150~\GeV\), is required to reflect the presence of \None and neutrinos in the signals events while rejecting a large part of the SM processes without neutrinos, particularly \Zjets events that include a fake lepton.
The fully leptonically decaying \WZ process is the leading SM background in \SRSFOS due to the single neutrino from the \Wboson decay.
This background is suppressed by requiring a large transverse mass \mT defined by:
\begin{equation*}
\mT = \sqrt{2\pT(\ell_W)\met(1-\cos(\Delta\phi(\ell_W,\Ptmiss)))},
\end{equation*}
where $\ell_W$ is the lepton candidate from the \Wboson decay defined by the left-over lepton when forming a SFOS lepton pair.
Since, in the \WZ events, \Ptmiss represents the \pT of the neutrino, the \mT distribution exhibits a kinematic edge at the \Wboson mass reflecting the \Wboson's Jacobian peak.
Multiple $\ell_W$ can be defined when the three leptons have the same flavor.
To account for such cases, the minimum \mT of all $\ell_W$ candidates (\mTmin) in \SRSFOS must satisfy \(\mTmin>125~\GeV\).

\begin{table}[t!]
\centering
\caption{\label{tab:SRdefinition} Summary of selections for signal region definition.
Merged cells indicate common selections.
The `--' indicates that no requirement is applied to the variable.
The number of same-flavor opposite-charge (SFOS) and different-flavor opposite-charge (DFOS) lepton pairs are represented by \(n_{\text{SFOS}}\) and \(n_{\text{DFOS}}\) respectively.
If more than one SFOS lepton pair is in the event, the invariant mass closest to the \Zboson-boson mass is quoted for \mll.
\SRSFOSon and \SRSFOSoff are divided into three \mlll bins and further divided into four channels by lepton flavor.
\SRSS is divided into three channels according to the lepton flavor.
}
\begin{threeparttable}[b]
\begin{tabular*}{\textwidth}{@{\extracolsep{\fill}}l|c|c|c|c|c|c|c|c|c|c|c}
\toprule
& \multicolumn{4}{c|}{} & \multicolumn{4}{c|}{} & \multicolumn{3}{c}{} \\
& \multicolumn{4}{c|}{\SRSFOSon} & \multicolumn{4}{c|}{\SRSFOSoff} & \multicolumn{3}{c}{\SRSS} \\
& \multicolumn{4}{c|}{} & \multicolumn{4}{c|}{} & \multicolumn{3}{c}{} \\
\multirow{2}{*}{Variables} & \multirow{2}{*}{\(eee\)} & \multirow{2}{*}{\(ee\mu\)} & \multirow{2}{*}{\(e\mu\mu\)} & \multirow{2}{*}{\(\mu\mu\mu\)} & \multirow{2}{*}{\(eee\)} & \multirow{2}{*}{\(ee\mu\)} & \multirow{2}{*}{\(e\mu\mu\)} & \multirow{2}{*}{\(\mu\mu\mu\)} & \multirow{2}{*}{\(eee\)} & \multirow{2}{*}{\(ee\mu\)} & \(2\mu\) \\
& & & & & & & & & & & \small{\((e\mu\mu+\mu\mu\mu)\)} \\
\midrule\midrule
Trigger & \multicolumn{11}{c}{single-lepton} \\
\(n_{\ell}^{\text{baseline}}\), \(n_{\ell}^{\text{signal}}\) & \multicolumn{11}{c}{\(=3\)} \\
\ptlepone, \ptleptwo, \ptlepthree[\GeV] & \multicolumn{11}{c}{\(> 28,\,20,\,10\)} \\
\(n_{\bjets}\) & \multicolumn{11}{c}{\(=0\)} \\
\midrule
\(n_{\text{SFOS}}\) & \multicolumn{4}{c|}{\(\geq 1\)} & \multicolumn{4}{c|}{\(\geq 1\)} & \multicolumn{3}{c}{\(= 0\)} \\
\(n_{\text{DFOS}}\) & \multicolumn{4}{c|}{--} & \multicolumn{4}{c|}{--} & \multicolumn{3}{c}{\(= 0\)} \\
\(\mll\) [\GeV] & \multicolumn{4}{c|}{\(\in[81.2, 101.2]\)} & \multicolumn{4}{c|}{\(\notin[81.2, 101.2]\)} & \multicolumn{3}{c}{--} \\
\(\mTmin\) [\GeV] & \multicolumn{4}{c|}{\(> 125\)} & \multicolumn{4}{c|}{\(> 125\)} & \multicolumn{3}{c}{--} \\
\(\met\) [\GeV] & \multicolumn{4}{c|}{\(> 150\)} & \multicolumn{4}{c|}{\(> 150\)} & \multicolumn{3}{c}{\(> 50\)} \\
\midrule\midrule
& \multicolumn{8}{c|}{\(\texttt{a}\) : \(\in[30, 200)\)} & \multicolumn{3}{c}{} \\
\mlll binning [\GeV]\tnote{a} & \multicolumn{8}{c|}{\(\texttt{b}\) : \(\in[200, 400)\)} & \multicolumn{3}{c}{--} \\
& \multicolumn{8}{c|}{\(\texttt{c}\) : \(\in[400, +\infty)\)} & \multicolumn{3}{c}{} \\
\bottomrule
\end{tabular*}
\begin{tablenotes}
\footnotesize
\item[a]{The \mlll binning applies separately to each flavor channel of \SRSFOS.}
\end{tablenotes}
\end{threeparttable}
\end{table}
\begin{figure}[t!]
\centering
\includegraphics[width=0.75\textwidth]{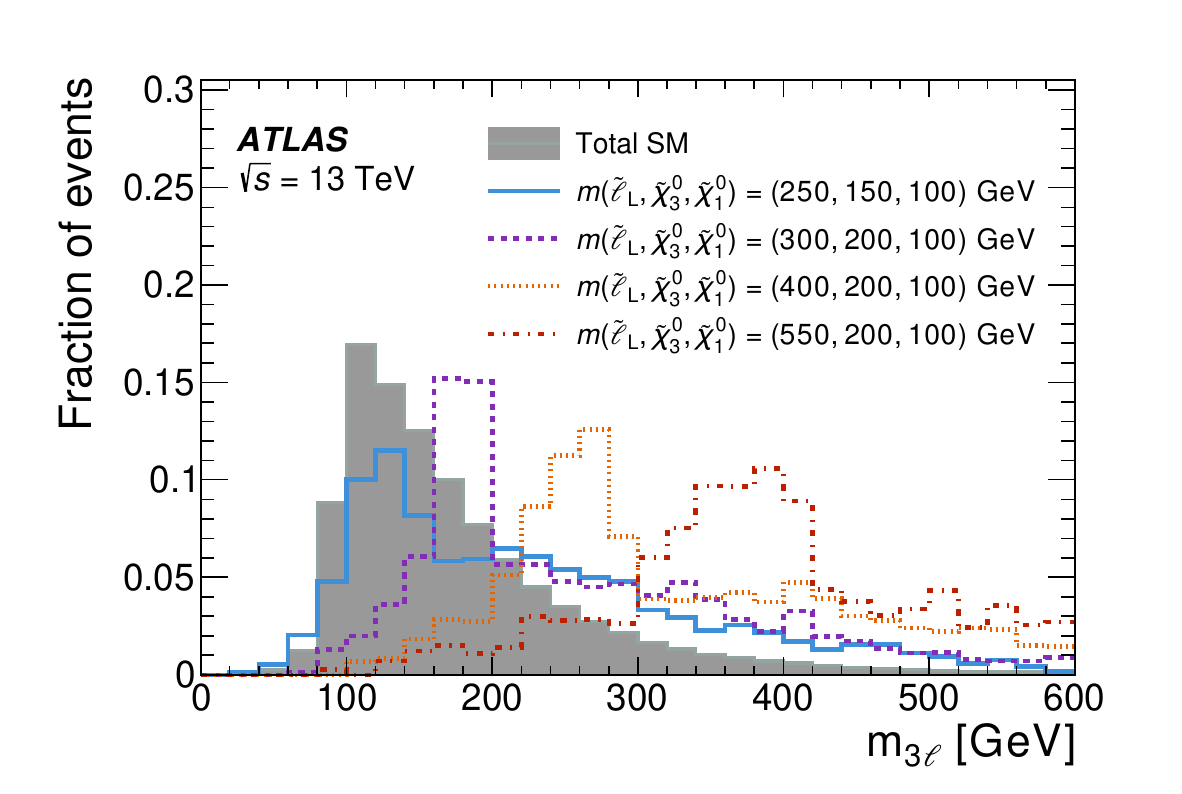}
\caption{Distribution of \mlll for the SM backgrounds and several signal hypotheses, normalized to unity.
All SM backgrounds are included and no uncertainties are shown.
A preselection is applied requesting exactly three baseline and signal leptons, and of at least one SFOS lepton pair.
}
\label{fig:SR3LSFOS_m3l}
\end{figure}
\begin{figure}[t!]
\centering
\includegraphics[width=0.90\textwidth]{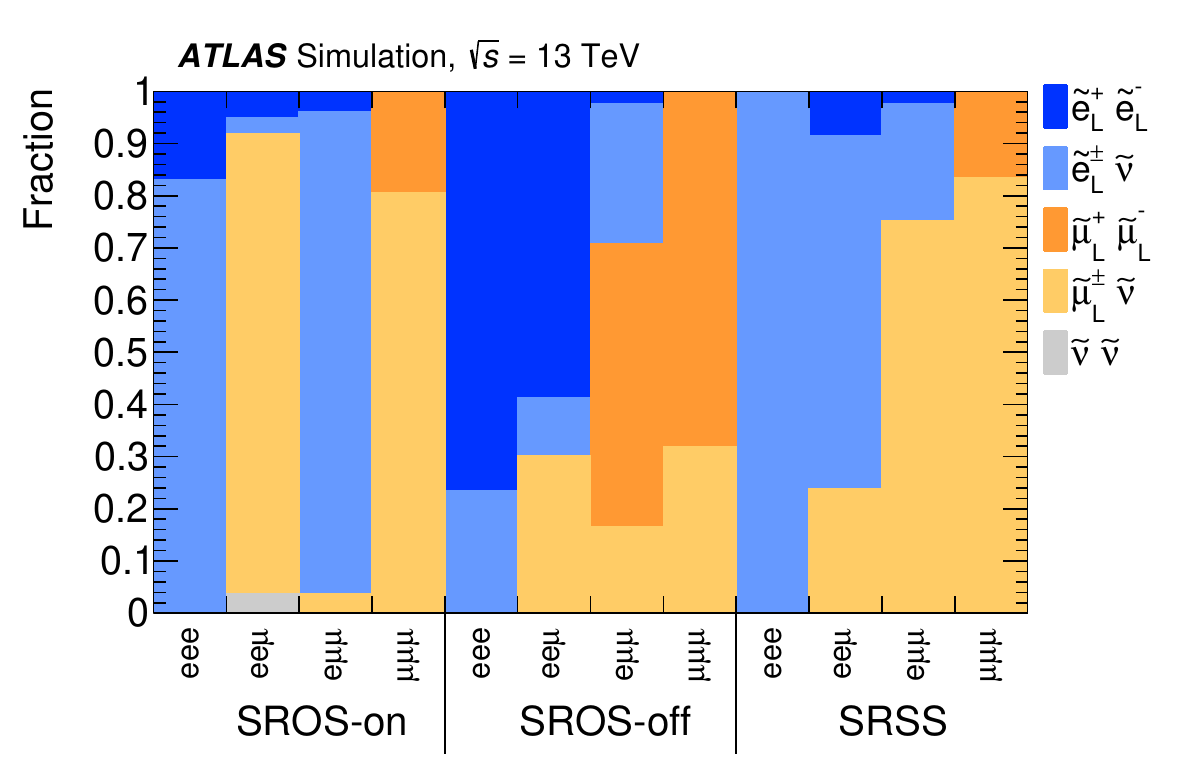}
\caption{Relative fractions of signal production modes populating the SR channels.
A signal with \(m(\sleptonL, \Nthree, \None)=(300, 200, 100)~\GeV\) is shown as an example.
Mass-degenerate \selL, \smuL and \snu is considered.
The \mlll bins are merged and \SRSSmm is divided into \SRSS-\(e\mu\mu\) and \SRSS-\(\mu\mu\mu\) for illustration.
}
\label{fig:signalclassify}
\end{figure}

A multi-bin strategy is applied in each \SRSFOS using the trilepton invariant mass, \mlll, and the lepton flavor combination to maximize the sensitivity across the model phase space.
First, each \SRSFOS is divided into \mlll bins, exploiting the cut-off structure of the signal as illustrated in Figure~\ref{fig:SR3LSFOS_m3l}.
The position of the cut-off corresponds to the mass splitting between the \sleptonL and \None, when the three leptons originate from the same side of the decay chain as shown in Figure~\ref{fig:Diagram}\subref{fig:slsnZV}.
In contrast, the distribution of backgrounds, dominated by \WZ events, smoothly falls without particular structures.
Each \SRSFOS is further divided into four channels by lepton flavor combination ($eee, ee\mu, e\mu\mu, \mu\mu\mu$) to maximize the sensitivity to the single-slepton-flavor models.

In the \SRSS, events are required to have three same-charge leptons.
The region with three same-charge leptons and zero \bjets has never been investigated before at the LHC.
The dominant SM backgrounds in this region are events with a charge-flip electron or a fake lepton.
The charge-flip-electron events majorly come from the \WZ process when the charge of one of the electrons from $Z \rightarrow ee$ is misidentified.
An \(\met>50~\GeV\) cut is imposed to exploit the real missing transverse momentum from the neutralinos and neutrinos in the targeted signals.
The \SRSS is further divided into three channels according to the lepton flavor combination, \(eee\), \(ee\mu\) and \(2\mu\), with the latter containing both the \(e\mu\mu\) and \(\mu\mu\mu\) channels to ease the background estimation due to extremely low background levels in these regions.

The breakdown of signal events per production mode in the various SR channels is summarized in Figure~\ref{fig:signalclassify}.
The selectron production signals tend to populate the \(eee\) and \(e\mu\mu\) channels in \SRSFOSon, and the \(eee\) and \(ee\mu\) channels in \SRSFOSoff and \SRSS.
Conversely, the smuon production signals are more likely to fall into the \(ee\mu\) and \(\mu\mu\mu\) channels in \SRSFOSon, and the \(e\mu\mu\) and \(\mu\mu\mu\) channels in \SRSFOSoff and \SRSS.
The selection acceptance times efficiency for \SRSFOSon, \SRSFOSoff and \SRSS are respectively \(5.4\times10^{-3}\), \(3.2\times10^{-3}\), and \(1.0\times10^{-3}\),
for a signal point with \(m(\sleptonL, \Nthree, \None)=(400, 200, 100)~\GeV\).
%


%

%
\section{Background estimation and validation}
\label{sec:background}

The SM backgrounds in this analysis are classified into two categories:
`irreducible backgrounds' where the three leptons are all real and promptly produced with the charge correctly assigned;
`reducible backgrounds' where one or more of the leptons are either a charge-flip or a fake lepton.
The irreducible backgrounds are particularly relevant in \SRSFOS accounting for 90\% of the total background, and are estimated using MC simulations corrected to data.
The reducible backgrounds account for almost all background in \SRSS.
These are estimated with dedicated data-driven approaches since the instrumental effects causing them are generally difficult to model.
The \ttbar background is exceptionally estimated with MC despite involving one fake lepton, since its MC modeling is found to be good, as shown in Section~\ref{subsec:fake}.

\subsection{Irreducible background estimation}
The dominant irreducible background process is  \WZ production.
A partially data-driven method is used for its estimation.
The \WZ MC is normalized to the data in a control region (CR), \CRWZ, designed with the same selection as \SRSFOS except for \(\mTmin\) being shifted to $40~\GeV \leq \mTmin < 80~\GeV$.
Dedicated validation regions (VRs) are set to validate the normalization and its extrapolation to the \SRSFOS.
These are defined either in the phase space between the \CRWZ and \SRSFOS ($80~\GeV \leq \mTmin < 125~\GeV$),
or in the low \met phase space with respect to the \SRSFOS ($\met<150~\GeV$).
The selections are summarized in Table~\ref{tab:CRVRWZ} and the region segmentation is illustrated in Figures~\ref{fig:SFOSonRegion}-\ref{fig:SFOSoffRegion}.
The \met range of the VRs is enlarged in the off-shell $Z$ region to supress signal contamination.
While the \CRWZ is inclusive in \mlll, the VRs follow the same splitting in \mlll as in the \SRSFOS.
The \WZ purity is about 94\% in \CRWZ and 54\%--93\% in VRs.
The signal contamination is at most 3.2\% in \CRWZ and 12\% in the VRs, with the largest signal contamination found in \VRWZoffmT-\texttt{c}.

\begin{table}[t!]
\centering
\caption{\label{tab:CRVRWZ} Summary of the selection criteria for the CR and VRs for \WZ, for the \SRSFOSon and \SRSFOSoff selection.
Merged cells indicate common selections.
The `--' indicates that no requirement is applied to the variable.
The number of same-flavor opposite-charge (SFOS) lepton pairs is represented by \(n_{\text{SFOS}}\).
If more than one SFOS lepton pair is in the event, the invariant mass closest to the \Zboson-boson mass is quoted for \mll.
VRs are divided into three \mlll bins to match the \SRSFOS binning.
}
\begin{tabular*}{\textwidth}{@{\extracolsep{\fill}}l|c|cc|cc}
\toprule
& & \multicolumn{2}{c|}{} & \multicolumn{2}{c}{} \\
& \CRWZ & \VRWZonmT & \VRWZonMET & \VRWZoffmT & \VRWZoffMET \\
Variables & & \multicolumn{2}{c|}{} & \multicolumn{2}{c}{} \\
\midrule\midrule
Trigger & \multicolumn{5}{c}{single-lepton} \\
\(n_{\ell}^{\text{baseline}}\), \(n_{\ell}^{\text{signal}}\) & \multicolumn{5}{c}{\(=3\)} \\
\ptlepone, \ptleptwo, \ptlepthree[\GeV] & \multicolumn{5}{c}{\(> 28,\,20,\,10\)} \\
\(n_{\bjets}\) & \multicolumn{5}{c}{\(=0\)} \\
\(n_{\text{SFOS}}\) & \multicolumn{5}{c}{\(\geq 1\)} \\
\midrule
\(\mll\) [\GeV] & -- & \multicolumn{2}{c|}{\(\in[81.2, 101.2]\)} & \multicolumn{2}{c}{\(\notin[81.2, 101.2]\)} \\
\(\mTmin\) [\GeV] & \(\in[40, 80)\) & \(\in[80, 125)\) & \(\in[125, +\infty)\) & \(\in[80, 125)\) & \(\in[125, +\infty)\) \\
\(\met\) [\GeV] & \(\in[150, +\infty)\) & \(\in[150, +\infty)\) & \(\in[80, 150)\) & \(\in[50, +\infty)\) & \(\in[20, 150)\) \\
\(n_{\text{jets}}\) & -- & -- & \(=0\) & -- & \(=0\) \\
\midrule\midrule
& & \multicolumn{4}{c}{\(\texttt{a}\) : \(\in[30, 200)\)} \\
\mlll binning [\GeV] & -- & \multicolumn{4}{c}{\(\texttt{b}\) : \(\in[200, 400)\)} \\
& & \multicolumn{4}{c}{\(\texttt{c}\) : \(\in[400, +\infty)\)} \\
\midrule\midrule
\WZ purity & 94\% & 93\% & 80\% & 87\% & 54\% \\
\bottomrule
\end{tabular*}
\end{table}

\begin{figure}[t!]
\centering
\subfloat[]{
\includegraphics[width=0.48\textwidth, keepaspectratio]{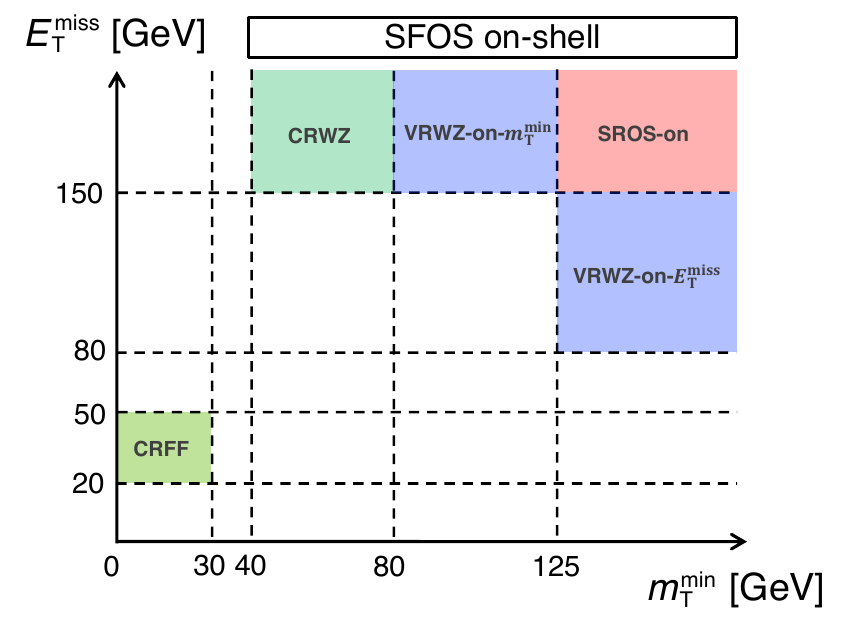}
\label{fig:SFOSonRegion}
}
\subfloat[]{
\includegraphics[width=0.48\textwidth, keepaspectratio]{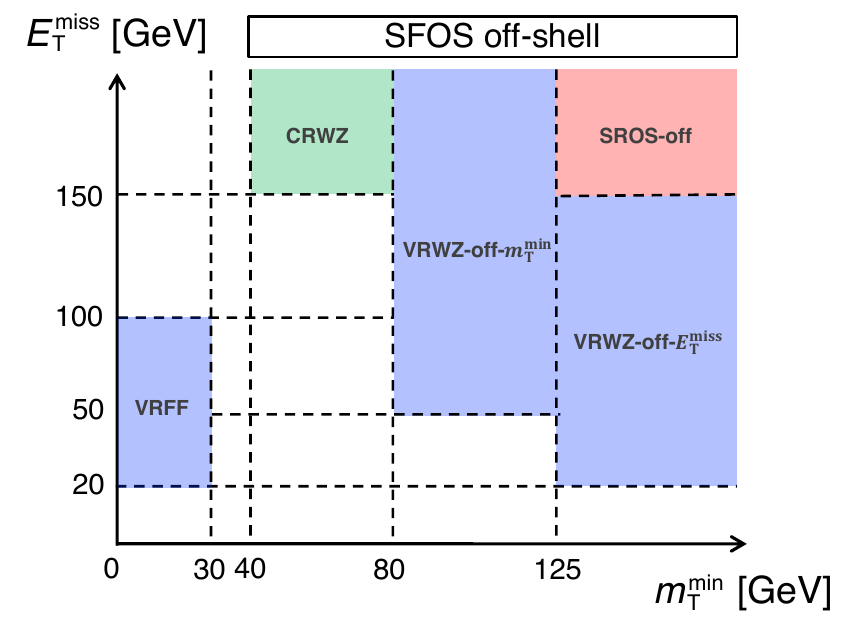}
\label{fig:SFOSoffRegion}
} \\
\subfloat[]{
\includegraphics[width=0.38\textwidth, keepaspectratio]{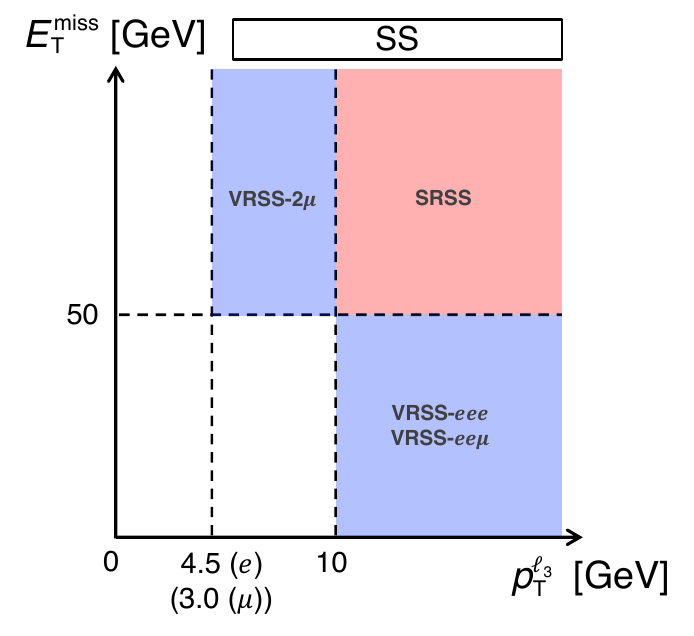}
\label{fig:SSRegion}
}
\subfloat[]{
\includegraphics[width=0.35\textwidth, keepaspectratio]{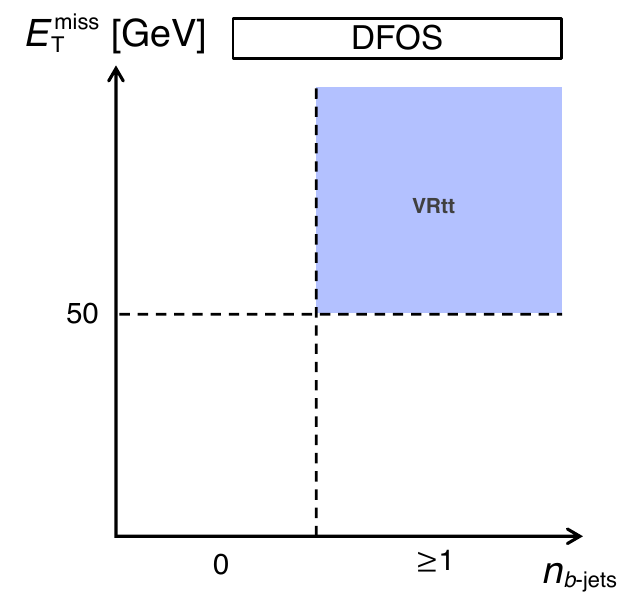}
\label{fig:DFOSRegion}
}
\caption{Schematics illustrating the selection of the CRs and the VRs used to estimate \protect\subref{fig:SFOSonRegion}\protect\subref{fig:SFOSoffRegion} the \WZ and fake background in \SRSFOS, \protect\subref{fig:SSRegion} the charge-flip and fake backgrounds in \SRSS, and \protect\subref{fig:DFOSRegion} the \ttbar background in \SRSFOS.
\label{fig:RegionSummary}}
\end{figure}

\begin{figure}[t!]
\centering
\includegraphics[width=0.96\textwidth]{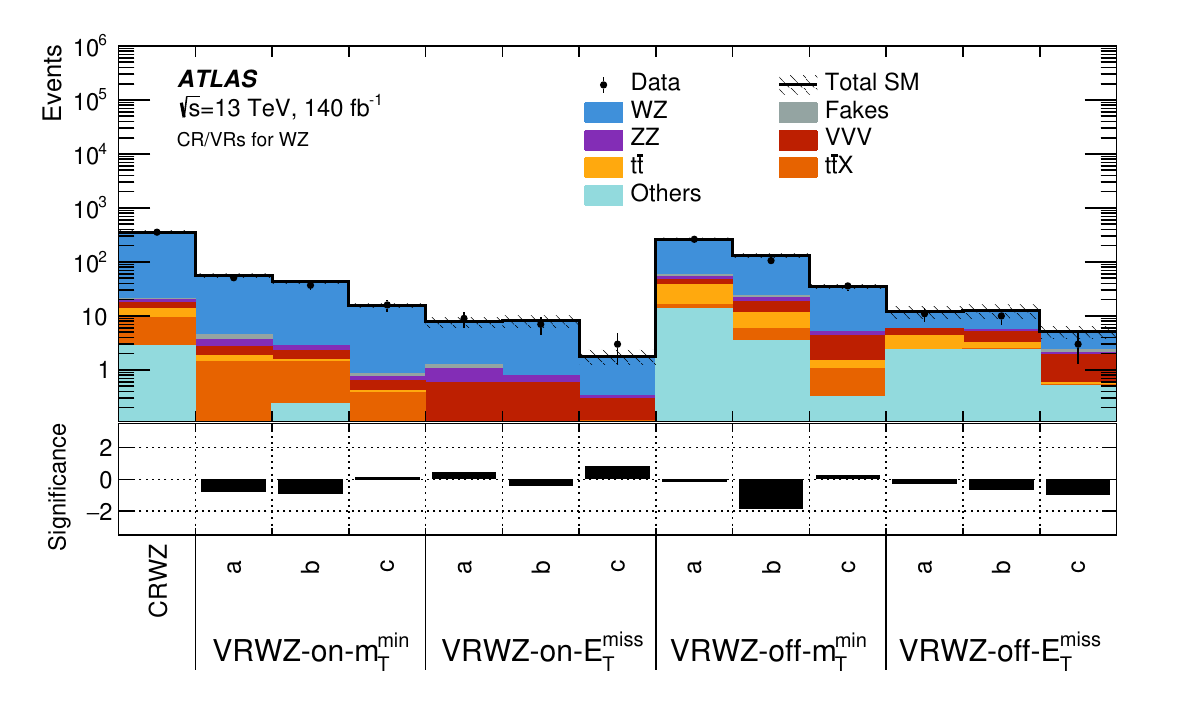}
\caption{Expected SM backgrounds and the data yields in the \CRWZ, \VRWZonmT, \VRWZonMET, \VRWZoffmT, and \VRWZoffMET designed to provide and validate the \WZ estimation in \SRSFOS.
The expected backgrounds are obtained from a background-only fit described in Section~\ref{subsec:StatAna}.
The `Others' category contains the production of Higgs boson, 3-top, 4-top, and single-top  processes.
The hatched band includes all statistical and systematic uncertainties discussed in Section~\ref{sec:systematics}.
The bottom panel shows the statistical significance~\cite{Cousins:2007bmb} of the difference between the observed events and the SM expectation.
}
\label{fig:histpull_Post-fit_WZ}
\end{figure}

\begin{figure}[t!]
\centering
\subfloat[]{
\includegraphics[width=0.49\textwidth, keepaspectratio]{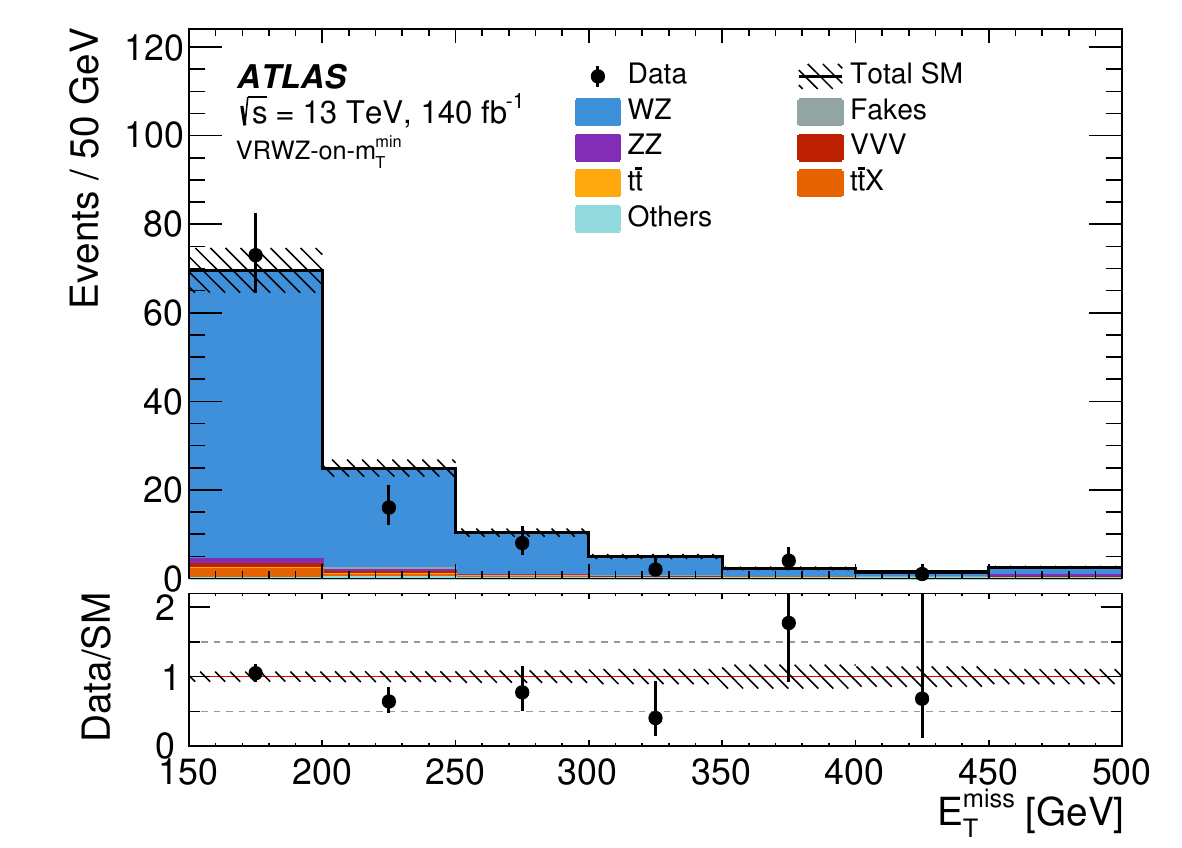} \label{fig:IrreducibleKinematicsVRWZon_lowmT_met_Et}
}
\subfloat[]{
\includegraphics[width=0.49\textwidth, keepaspectratio]{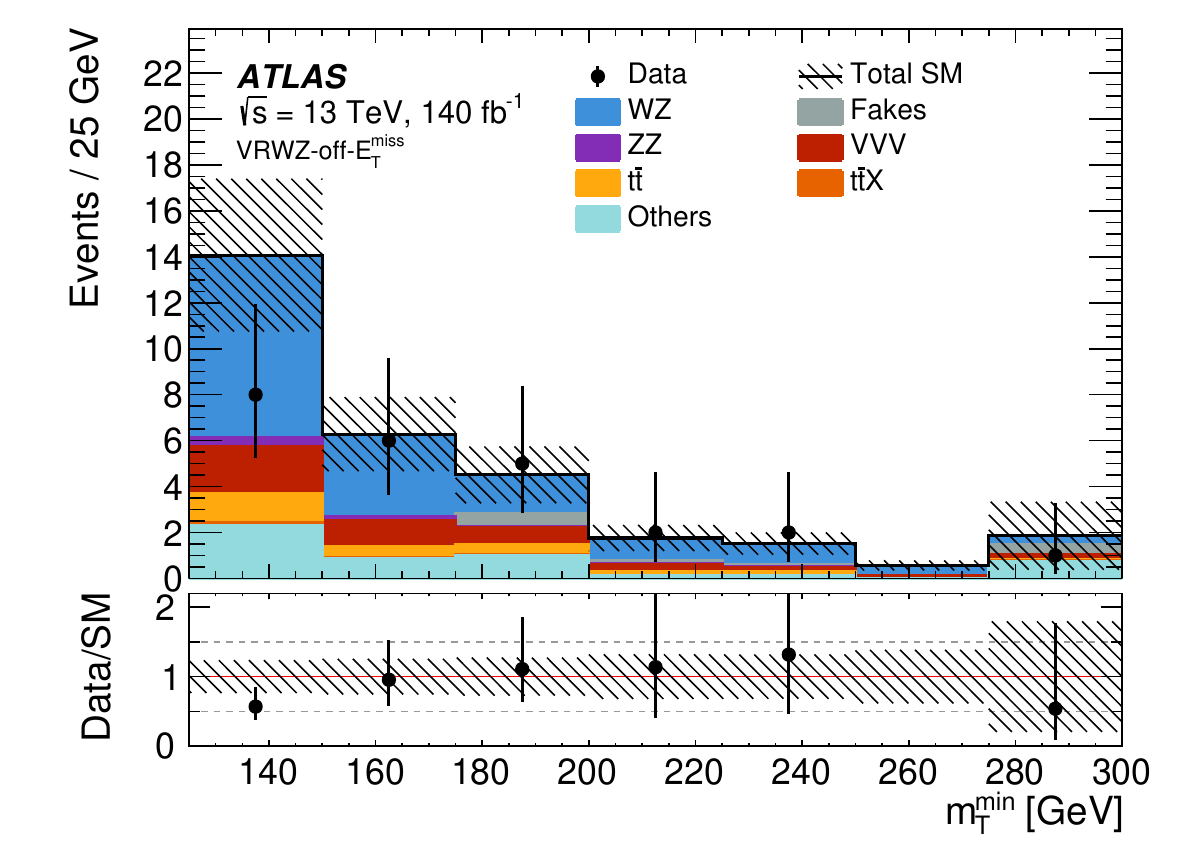} \label{fig:IrreducibleKinematicsVRWZoff_lowMET_mTSFOS_min}
}
\caption{Example kinematic distributions for the expected backgrounds and the data,
obtained from a background-only fit described in Section~\ref{subsec:StatAna}.
The figure shows \protect\subref{fig:IrreducibleKinematicsVRWZon_lowmT_met_Et} the \met distribution in \VRWZonmT and \protect\subref{fig:IrreducibleKinematicsVRWZoff_lowMET_mTSFOS_min} the \mTmin distribution in \VRWZoffMET.
The last bin includes the overflow.
The `Others' category contains the production of Higgs boson, 3-top, 4-top, and single-top  processes.
The hatched band includes all statistical and systematic uncertainties discussed in Section~\ref{sec:systematics}.
The bottom panel shows the ratio of the observed data to the predicted yields.
\label{fig:IrreducibleKinematics}}
\end{figure}

The other rare irreducible SM processes, including triboson ($VVV$), Higgs boson production, top-pair production in association with a boson ($\ttbar X$), 3-top, 4-top, and single-top production, are estimated from MC simulations with SM cross-sections in all analysis regions.

The \WZ normalization factor extracted from \CRWZ is found to be \(1.07\pm0.06\).
The estimated background and the observed data in the relevant VRs are shown in Figure~\ref{fig:histpull_Post-fit_WZ},
and some example kinematic distributions in the VRs are presented in Figure~\ref{fig:IrreducibleKinematics}.
Good agreement is generally observed.

\subsection{Charge-flip background estimation}
\label{subsec:chargeflip}
Charge-flip leptons are predominantly electrons which emit bremmstrahlung while propagating through the detector material~\cite{EXOT-2018-34,EGAM-2018-01}, creating electron-positron pairs.
The production of these secondary particles can lead to distortions of the primary electron track, when hits from secondary particles are included in the fit, or when the track from a secondary particle is wrongly selected as the primary track.

The charge-flip background is only relevant in \SRSS.
The MC modeling of charge flip is not always reliable since they are sensitive to details of the detector modeling.
In this analysis, the per-electron charge-flip probability is measured using \Zee data and a MC correction factor (`scale factor') is derived as a function of the \pT and \abseta of the electron.
The scaled MC simulation is then used to estimate the charge-flip background in \SRSS.

The charge-flip probability measurement in the \Zee data is based on a likelihood fit as described in Ref.~\cite{EGAM-2018-01}.
The measured probability is then compared with the \Zee MC simulation to derive the scale factors.
Typically the charge-flip probability is of the order of \(\mathcal{O}(10^{-5})\) in the low-\pT region,
increases with increasing \pT and \abseta, and may reach as high as \(\mathcal{O}(0.1)\).
High-\pT electrons are more likely to cause charge-flip, as their tracks are approximately straight and more susceptible to small angular perturbations.
The higher charge-flip probability in high-\abseta electrons is due to denser detector materials in this region.
Systematic uncertainties in the charge-flip scale factors are assigned based on the statistical uncertainties of the data sample,
modeling uncertainties in the background subtraction,
and uncertainties related to the parameterization of charge-flip probabilities.
Examples of the measured charge-flip probabilities and the scale factors are illustrated in Figure~\ref{fig:ChargeFliprate}.

\begin{figure}[t!]
\centering
\subfloat[]{
\includegraphics[width=0.48\textwidth, keepaspectratio]{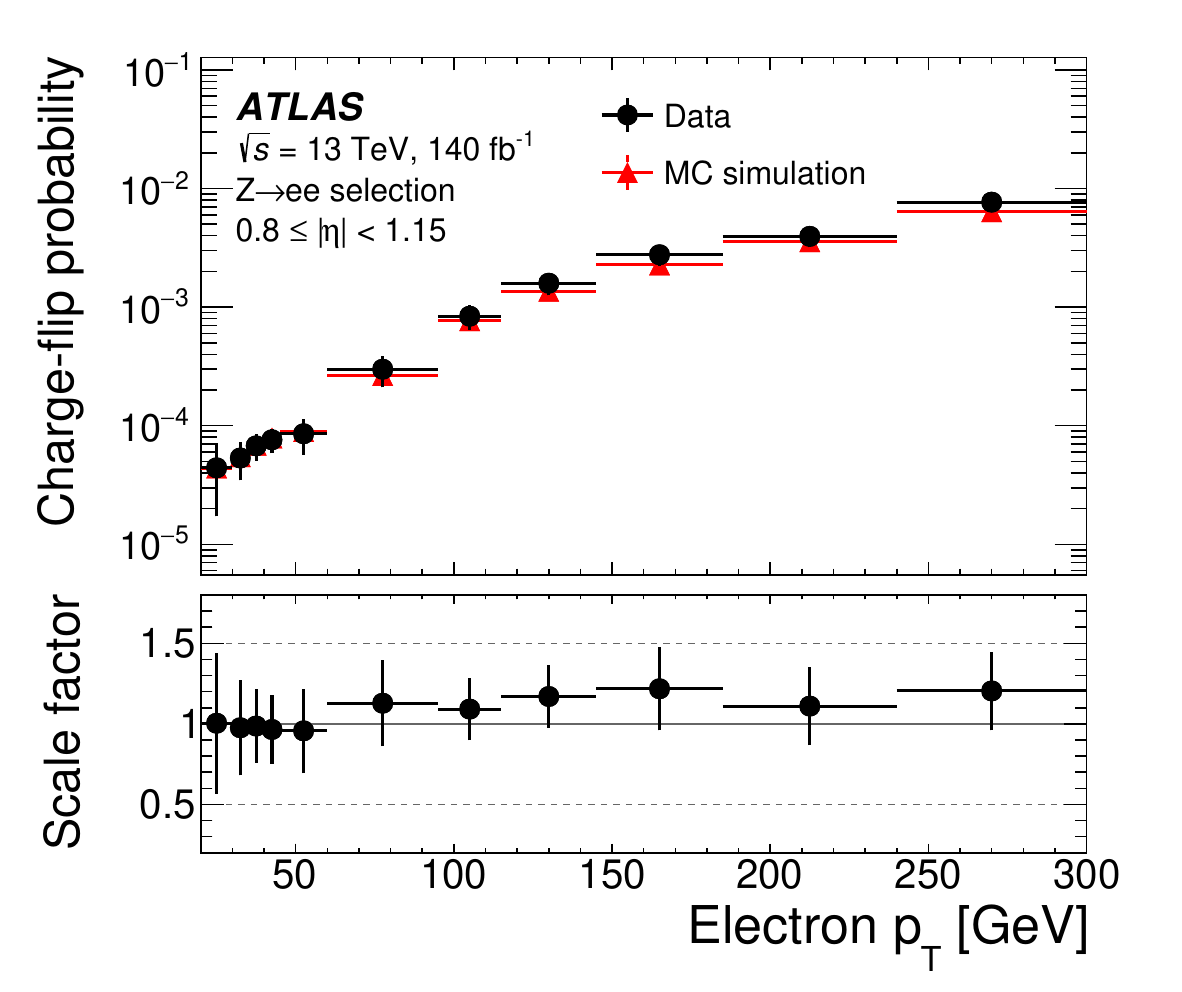} \label{fig:ChargeFliprate_eta}
}
\subfloat[]{
\includegraphics[width=0.48\textwidth, keepaspectratio]{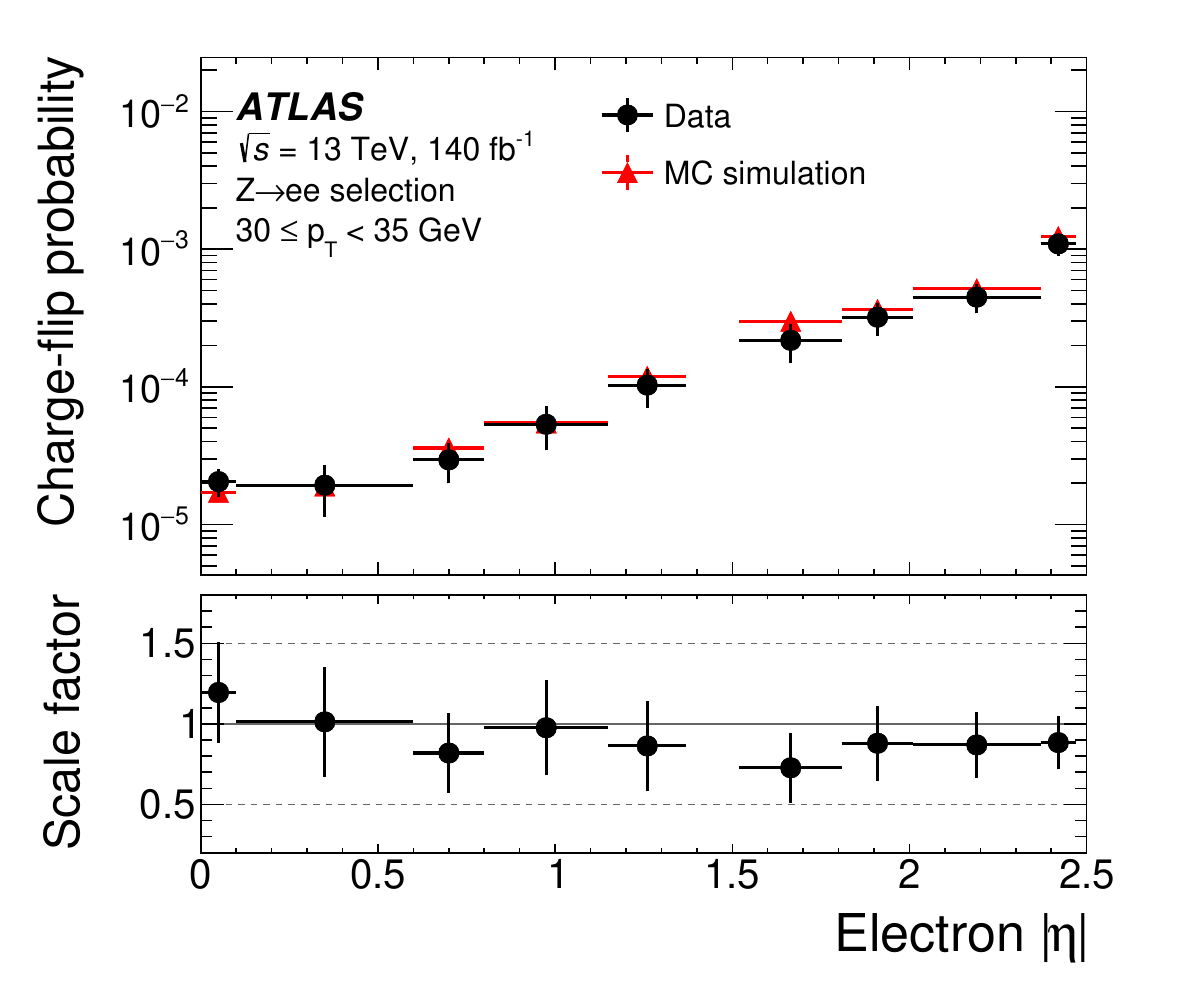} \label{fig:ChargeFliprate_pT}
}
\caption{Charge-flip (upper panel) probabilities in data and MC simulation and (lower panel) scale factors \protect\subref{fig:ChargeFliprate_eta} as a function of the electron \pT for a specific electron \abseta slice (\(0.8 \leq \abseta < 1.15\)) or \protect\subref{fig:ChargeFliprate_pT} as a function of the electron \abseta for a specific electron \pT slice (\(30 \leq \pT < 35~\GeV\)). For \protect\subref{fig:ChargeFliprate_eta}, the highest-\pt bin includes the overflow.
The uncertainties in the data charge-flip probabilities and the scale factors include both the statistical and systematic components, while the uncertainties in MC simulation charge-flip probabilities include only the statistical uncertainty. \label{fig:ChargeFliprate}}
\end{figure}

An additional uncertainty is assigned to account for the physics process dependency of the charge-flip probabilities when applying scale factors in the analysis.
While the scale factors are measured using \Zee events, the main charge-flip source in \SRSS is \WZ.
Though charge-flip probabilities are found to be statistically consistent between \Zee and \WZ, to ensure the difference is covered, a 20\% systematic uncertainty is assigned to account for possible physics-process dependency.
This is evaluated based on a MC closure test, comparing the MC yields of same-sign \WZ events to those of reweighted opposite-sign \WZ events, using MC-derived charge-flip probabilities of the \Zee process.

The validation regions \VRSSeee and \VRSSeem are defined to test the charge-flip estimation in \SRSS,
by inverting the \met selection with respect to \SRSSeee and \SRSSeem, respectively.
To further boost the data event statistics, a set of supplementary validation regions, \VRSSnoECIDS, are defined by removing the ECIDS requirement to the leading lepton.
The selection is summarized in Table~\ref{tab:CRVRFFttSSdefinition} and the region segmentation is illustrated in Figure~\ref{fig:RegionSummary}\subref{fig:SSRegion}.
The signal contamination in \VRSSeee and \VRSSeem is below 8\% of the total expected background.
The observed and expected event yields for these VRs are summarized in Figure~\ref{fig:histpull_Post-fit_reducible}
with an example kinematic distribution shown in Figure~\ref{fig:ReducibleKinematics}~\subref{fig:ReducibleKinematics_CF}.
Reasonable agreement is observed.

\begin{table}[t!]
\centering
\caption{\label{tab:CRVRFFttSSdefinition} Definitions of the CR and VRs involved in the fake and charge-flip lepton estimation.
Merged cells indicate common selections.
The `--' indicates that no requirement is applied to the variable.
The number of same-flavor opposite-charge (SFOS) and different-flavor opposite-charge (DFOS) lepton pairs are represented by \(n_{\text{SFOS}}\) and \(n_{\text{DFOS}}\) respectively.
If more than one SFOS lepton pair is in the event, the invariant mass closest to the \Zboson-boson mass is quoted for \mll.}
\begin{tabular*}{1.0\textwidth}{@{\extracolsep{\fill}}l|c|c|c|c|c|c|c|c|c|c}
\toprule
& & \multicolumn{4}{c|}{} & \multicolumn{2}{c|}{} & \multicolumn{2}{c|}{} & \\
& \multirow{2}{*}{\CRFF} & \multicolumn{4}{c|}{\multirow{2}{*}{\VRFF}} & \multicolumn{2}{c|}{\multirow{2}{*}{\VRtt}} & \multicolumn{2}{c|}{\VRSS}& \multirow{2}{*}{\VRSS} \\
& & \multicolumn{4}{c|}{} & \multicolumn{2}{c|}{} & \multicolumn{2}{c|}{(\texttt{-noECIDS})} & \\
& & \multicolumn{4}{c|}{} & \multicolumn{2}{c|}{} & \multicolumn{2}{c|}{} & \\
\multirow{2}{*}{Variables} & \multirow{2}{*}{} & \multirow{2}{*}{\(eee\)} & \multirow{2}{*}{\(ee\mu\)} & \multirow{2}{*}{\(e\mu\mu\)} & \multirow{2}{*}{\(\mu\mu\mu\)} & \multirow{2}{*}{\(ee\mu\)} & \multirow{2}{*}{\(e\mu\mu\)} & \multirow{2}{*}{\(eee\)} & \multirow{2}{*}{\(ee\mu\)} & \(2\mu\) \\
& & & & & & & & & & \small{\((e\mu\mu+\mu\mu\mu)\)} \\
\midrule\midrule
Trigger & \multicolumn{10}{c}{single-lepton} \\
\(n_{\ell}^{\text{baseline}}\), \(n_{\ell}^{\text{signal}}\) & \multicolumn{10}{c}{\(=3\)} \\
\ptlepone, \ptleptwo[\GeV] & \multicolumn{5}{c|}{\(> 10,\,10\)} & \multicolumn{5}{c}{\(> 28,\,20\)} \\
\ptlepthree[\GeV] & \multicolumn{5}{c|}{\(> 10\)} & \multicolumn{2}{c|}{\(> 10\)} & \multicolumn{2}{c|}{\(> 10\)} & \(< 10\) \\
\(n_{\bjets}\) & \multicolumn{5}{c|}{\(=0\)} &\multicolumn{2}{c|}{\(\geq 1\)} & \multicolumn{2}{c|}{\(=0\)} & \(=0\) \\
\midrule
\(n_{\text{SFOS}}\) & \(\geq 1\) & \multicolumn{4}{c|}{\(\geq 1\)} & \multicolumn{2}{c|}{\(= 0\)} & \multicolumn{2}{c|}{\(= 0\)} & \(= 0\) \\
\(n_{\text{DFOS}}\) & -- & \multicolumn{4}{c|}{--} & \multicolumn{2}{c|}{\(\geq 1\)} & \multicolumn{2}{c|}{\(= 0\)} & \(= 0\) \\
\mll[\GeV] & \(\in[81.2, 101.2]\) & \multicolumn{4}{c|}{\(\notin[81.2, 101.2]\)} & \multicolumn{2}{c|}{--} & \multicolumn{2}{c|}{--} & -- \\
\mTmin[\GeV] & \(< 30\) & \multicolumn{4}{c|}{\(< 30\)} & \multicolumn{2}{c|}{--} & \multicolumn{2}{c|}{--} & -- \\
\met[\GeV] & \(\in [20, 50)\) & \multicolumn{4}{c|}{\(\in [20, 100)\)} & \multicolumn{2}{c|}{\(> 50\)} & \multicolumn{2}{c|}{\(< 50\)} & \(> 50\) \\
\mlll[\GeV] & \(> 105\) & \multicolumn{4}{c|}{\(> 105\)} & \multicolumn{2}{c|}{--} & \multicolumn{2}{c|}{--} & -- \\
\bottomrule
\end{tabular*}
\end{table}

\begin{figure}[t!]
\centering
\includegraphics[width=0.96\textwidth]{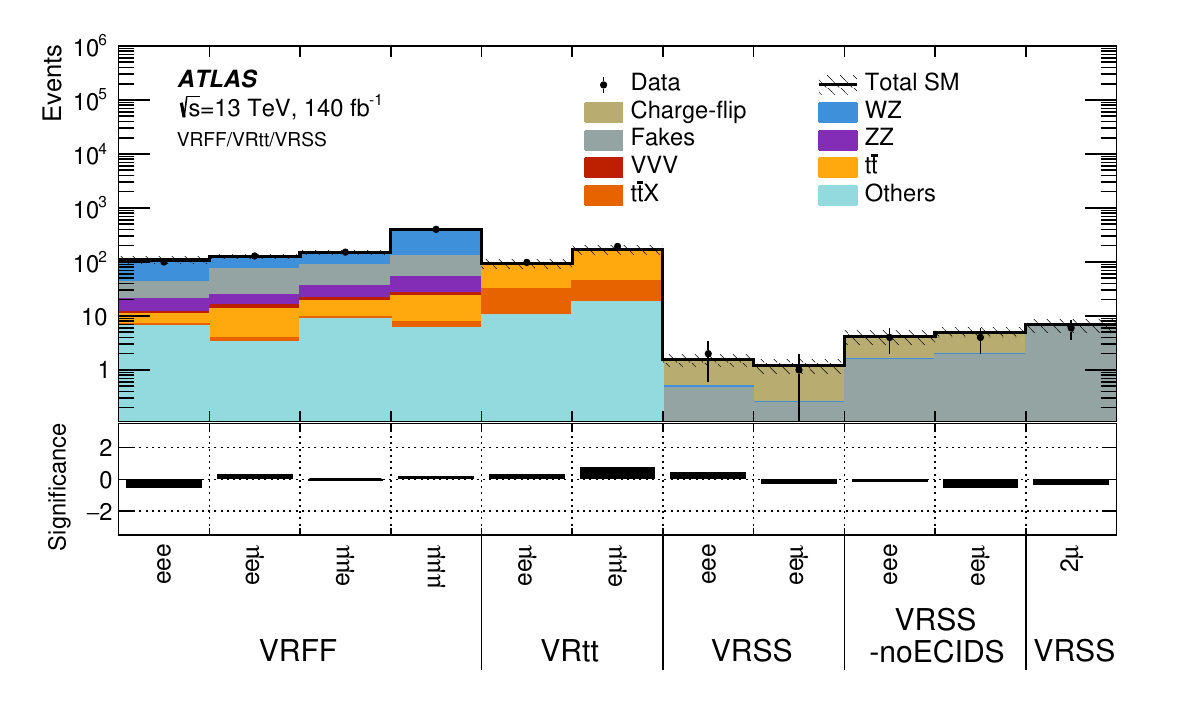}
\caption{Expected SM backgrounds and the data yields in the \VRFF, \VRtt, and \VRSS.
The expected backgrounds are obtained from a background-only fit described in Section~\ref{subsec:StatAna}.
The `Others' category contains the production of Higgs boson, 3-top, 4-top, and single-top  processes.
The hatched band includes all statistical and systematic uncertainties discussed in Section~\ref{sec:systematics}.
The bottom panel shows the statistical significance~\cite{Cousins:2007bmb} of the difference between the observed events and the SM expectation.
}
\label{fig:histpull_Post-fit_reducible}
\end{figure}
\begin{figure}[t!]
\centering
\subfloat[]{
\includegraphics[width=0.49\textwidth, keepaspectratio]{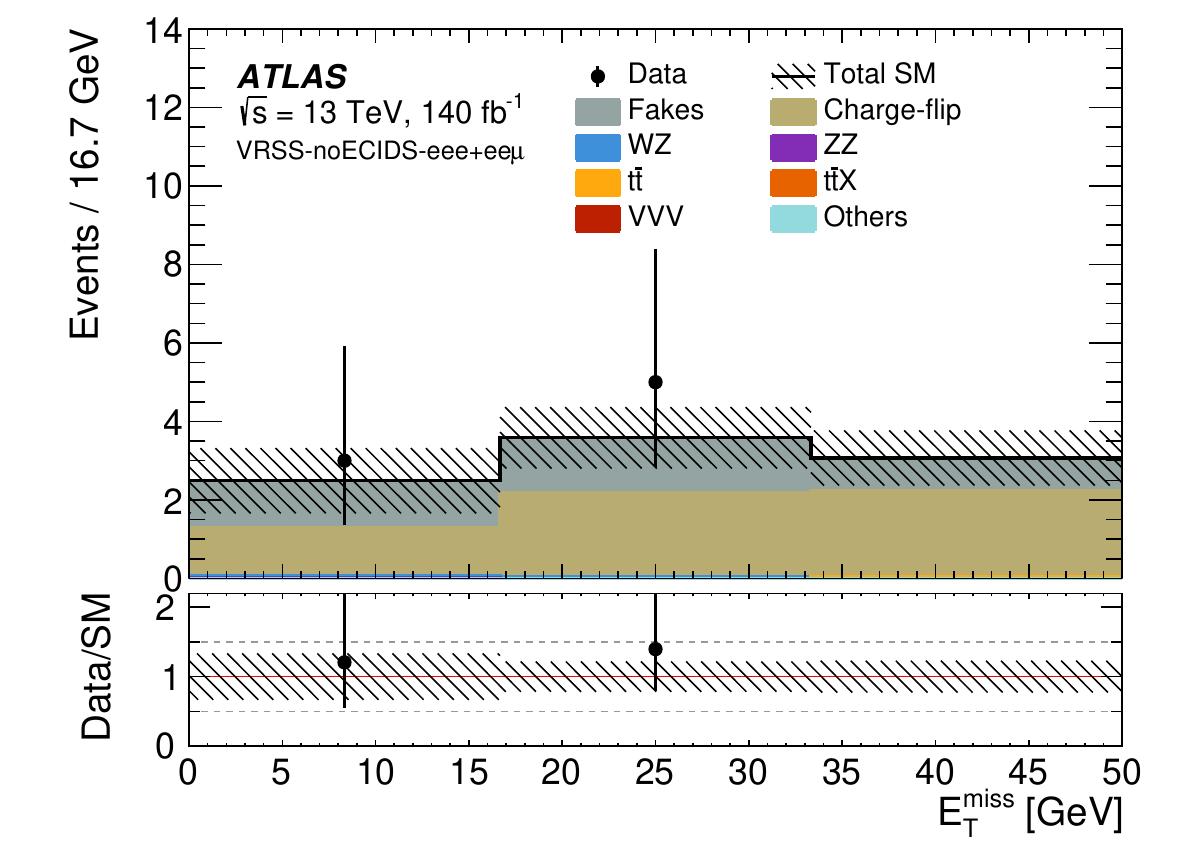} \label{fig:ReducibleKinematics_CF}
}
\subfloat[]{
\includegraphics[width=0.49\textwidth, keepaspectratio]{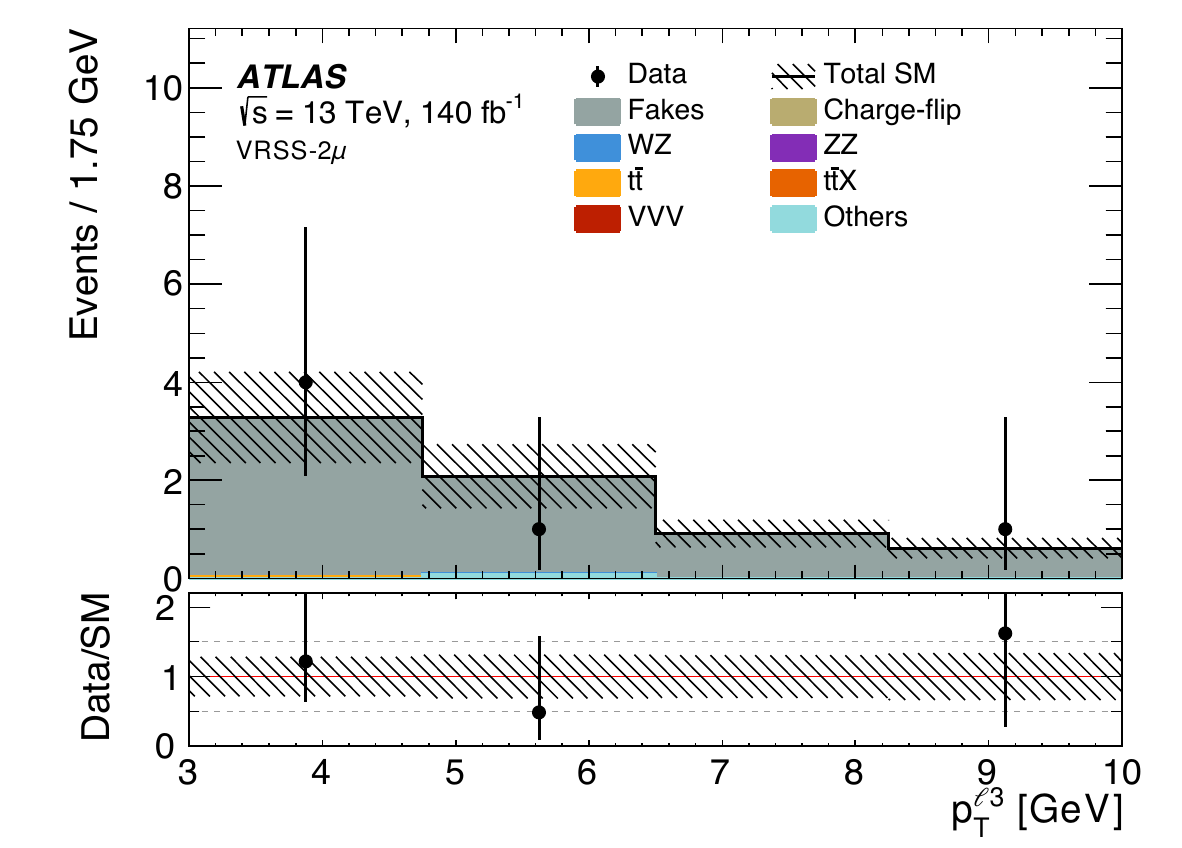} \label{fig:ReducibleKinematics_Fake}
}
\caption{Example kinematic distributions for the expected backgrounds and the data,
obtained from a background-only fit described in Section~\ref{subsec:StatAna}.
The figure shows \protect\subref{fig:ReducibleKinematics_CF} the \met distribution in the merged region of \VRSSnoECIDS-\(eee\) and \VRSSnoECIDS-\(ee\mu\) and \protect\subref{fig:ReducibleKinematics_Fake} the \ptlepthree distribution in \VRSSmm.
The `Others' category contains the production of Higgs boson, 3-top, 4-top, and single-top  processes.
The hatched band includes all statistical and systematic uncertainties discussed in Section~\ref{sec:systematics}.
The bottom panel shows the ratio of the observed data to the predicted yields.
\label{fig:ReducibleKinematics}}
\end{figure}

\subsection{Fake-lepton background estimation}
\label{subsec:fake}
Fake leptons are defined as either hadrons misidentified as leptons, or non-prompt real leptons from photon conversion or \(b\)- or \(c\)-hadron decays.
Electrons from final-state-radiation or bremsstrahlung photons originating from a prompt electron are not considered as fake leptons.\footnote{Events with such electrons are not targeted by the fake-factor method but are instead taken directly from MC simulation, which is considered to adequately model such processes.
These events have a minor contribution in \SRSS, and are negligible in all other regions.}
A data-driven method, referred to as the `fake-factor method'~\cite{STDM-2011-24,ATL-PHYS-PUB-2010-005}, is used to estimate the fake-lepton backgrounds.
A lepton `anti-ID' requirement is defined for the fake-factor method, corresponding to leptons that satisfy the baseline criteria but not the signal criteria.
The fake factor (FF) is defined as the ratio of the probability that a given lepton candidate satisfies the signal lepton requirement to the probability that it fulfills the anti-ID requirement.

The FF is measured using data in a control region, \CRFF, designed to be enriched with \Zjets events associated with a fake lepton.
The selection is summarized in Table~\ref{tab:CRVRFFttSSdefinition} and the region segmentation is illustrated in Figure~\ref{fig:RegionSummary}\subref{fig:SFOSonRegion}.
After selecting exactly three baseline leptons with at least one SFOS lepton pair,
the \Zboson-boson candidate in the event is identified as the SFOS pair yielding the invariant mass closest to the \Zboson-boson mass, and the remaining lepton is tagged as the fake-lepton candidate.
Either of the two leptons from the \Zboson-boson candidate must activate the single-lepton trigger to remove the bias from trigger selection on the fake leptons.
To suppress the \WZ contribution in \CRFF, the \mTmin and \met selections are imposed.
Additionally, a requirement of \(\mlll>105~\GeV\) is applied to suppress \(Z\rightarrow\ell\ell\gamma^{\ast}\rightarrow4\ell\).
The FF is extracted as the fraction of fake-lepton candidates satisfying to that failing the signal lepton criteria.

The fake-lepton background estimation in a given region is then obtained by applying the FF to the data events in its corresponding `anti-ID region', defined by requiring at least one anti-ID lepton as opposed to three signal leptons in the target region.
An exceptional treatment is applied to the estimation in \SRSSmm, where the FF is applied to the MC simulated events in the anti-ID region with two signal and one anti-ID leptons, instead of the data events.
This is due to the small data statistics in \SRSSmm anti-ID region, which would result in a large uncertainty.
This exception is validated by verifying the MC modeling in the anti-ID region with a loosened kinematic selection, and confirming the similar origin composition of fake leptons in \SRSSmm and its anti-ID region using MC simulation.

The FFs are derived separately per lepton flavor, and are parameterized as a function of lepton \pT.
Typical FF values are 0.04--0.08 (0.08--0.12) for electrons (muons) in a \pT range of 4.5--50 (3.0--30)~\GeV.
FFs are parameterized inclusively in lepton's $\eta$ since no significant $\eta$ dependency is found.
In both the FF measurement and the FF application procedure, contributions from \ttbar and irreducible background processes are subtracted using MC simulation samples.

The data-driven fake estimation is subject to the statistical uncertainties of the data in the anti-ID regions to which the FFs are applied,
and the uncertainties in the FF values.
The FF uncertainties include the statistical uncertainty of the data in \CRFF, an uncertainty obtained by varying the choice of the \met requirement in \CRFF to account for the potential fake-lepton composition change in data,
and an uncertainty in the subtraction of the \WZ process in \CRFF evaluated by varying the \WZ normalization by 10\%~\cite{STDM-2018-03}.
These FF uncertainties are propagated by varying the FF in the estimation.
An additional systematic uncertainty is considered to account for the difference in the fake-lepton origin compositions between the \CRFF and the SRs.
This is evaluated based on a MC closure test, comparing the MC prediction in a SR to the FF method estimation using the MC-derived FFs.

The fake-lepton estimates derived for \SRSFOS and \SRSS are validated in \VRFF and \VRSSmm, respectively.
The selection is summarized in Table~\ref{tab:CRVRFFttSSdefinition} and the region segmentation is shown in Figure~\ref{fig:RegionSummary}\subref{fig:SFOSoffRegion} and Figure~\ref{fig:RegionSummary}\subref{fig:SSRegion}.
The \VRFF is defined in the off-shell $Z$ region to enhance the fake-lepton fraction, and is separated by lepton flavor.
The purity of the fake-lepton background in \VRFF is about 25\%, while the contamination from signals is negligible.
The \VRSSmm is defined by inverting the selection on the third leading lepton \pT ($\pT<10~\GeV$) with respect to \SRSSmm.
The signal contamination in \VRSSmm is at most 15\%.
The observed and expected event yields for these VRs are summarized in Figure~\ref{fig:histpull_Post-fit_reducible}
with an example kinematic distribution shown in Figure~\ref{fig:ReducibleKinematics}~\subref{fig:ReducibleKinematics_Fake}.
Reasonable agreement is observed.

The \ttbar MC modeling is validated in \VRtt, enhancing the \ttbar contribution by requiring a different-flavor opposite-charge (DFOS) lepton pair, moderate \(\met\), and the presence of one or more \bjets.
The \ttbar purity is about 70\% in \VRtt.
Selection requirements for \VRtt are summarized in Table~\ref{tab:CRVRFFttSSdefinition} and illustrated in Figure~\ref{fig:DFOSRegion}.
The observed and expected event yields for \VRFF are summarized in Figure~\ref{fig:histpull_Post-fit_reducible}.


%

%
\section{Systematic uncertainties}
\label{sec:systematics}

Systematic uncertainties in the predicted event yields are considered for both the signal and background processes, and propagated to the final results through the profile likelihood described in Section~\ref{subsec:StatAna}.
Several sources of systematic uncertainty are considered.
They are grouped into experimental uncertainties, theoretical uncertainties, MC statistical uncertainties, \WZ normalization uncertainty, and uncertainties from the data-driven methods applied in this analysis.
The contribution to the analysis sensitivity is generally minor, since the total uncertainty in the SRs is dominated by the statistical uncertainty.

The experimental uncertainties encompass possible differences between data and simulations in all analysis elements including trigger, pileup, and reconstructed objects.
The leading experimental uncertainty originates from the electron identification efficiencies in the electron-dominated SRs, and the jet energy resolution in the muon-dominated SRs.
For leptons, uncertainties in the reconstruction efficiencies~\cite{EGAM-2018-01}, identification efficiencies~\cite{MUON-2018-03}, isolation efficiencies, energy scales~\cite{EGAM-2018-01,EGAM-2021-02}, resolutions, and trigger efficiencies are considered.
The uncertainties related to electrons typically yield a relative uncertainty in the total expected background of 10\% for the electron-dominated \SRSFOS.
For jets, uncertainties in the vertex tagger efficiency~\cite{PERF-2014-03}, %
energy scale~\cite{PERF-2016-04},
energy resolution~\cite{JETM-2018-05},
and efficiencies of the flavor tagging~\cite{FTAG-2018-01,FTAG-2020-08,FTAG-2019-02} are considered.
The uncertainties related to jets typically yield a relative uncertainty in the total expected background of 5\% for the muon-dominated \SRSFOS.
Uncertainties associated with the objects used to compute the \met are propagated through the computation,
and additional uncertainties in the scale and resolution of the soft term are also included~\cite{JETM-2020-03}.
The uncertainty in the combined 2015--2018 integrated luminosity is 0.83\%~\cite{DAPR-2021-01}, obtained primarily using the luminosity measurements of the LUCID-2 detector~\cite{LUCID2}.

The theoretical uncertainties comprise the cross-section uncertainty and the MC generator modeling uncertainty.
Signals and the background processes that are not normalized in the CR are subject to cross-section uncertainties.
For the triboson background, %
a \(+45/-40\%\) uncertainty is applied to \(WWW\) process, a \(+35/-30\%\) uncertainty to \(WWZ\), a \(+192/-100\%\) uncertainty to \(WZZ\), and a \(+440/-100\%\) uncertainty to \(ZZZ\), based on their cross-section measurement by the CMS Collaboration~\cite{CMS-SMP-19-014}\footnote{The ATLAS Collaboration has also produced compatible results~\cite{STDM-2020-08} in which the \(ZZZ\) process is however not separately measured.}.
Cross-section uncertainties of 13\%, 12\%, and 10\% are applied for the minor backgrounds \ttW, \ttZ, and \ttH, respectively~\cite{deFlorian:2016spz,TOPQ-2019-30,TOPQ-2020-20,HIGG-2021-23},
and a conservative uncertainty of 50\% is applied to all other rare top processes, which has little impact on final results.
For the signal samples, theoretical uncertainties in the cross-section are applied, ranging from 2\% at 120~\GeV to 6\% at 700~\GeV, as detailed in Section~\ref{sec:samples}.
\begin{figure}[t!]
\centering
\includegraphics[width=0.98\textwidth]{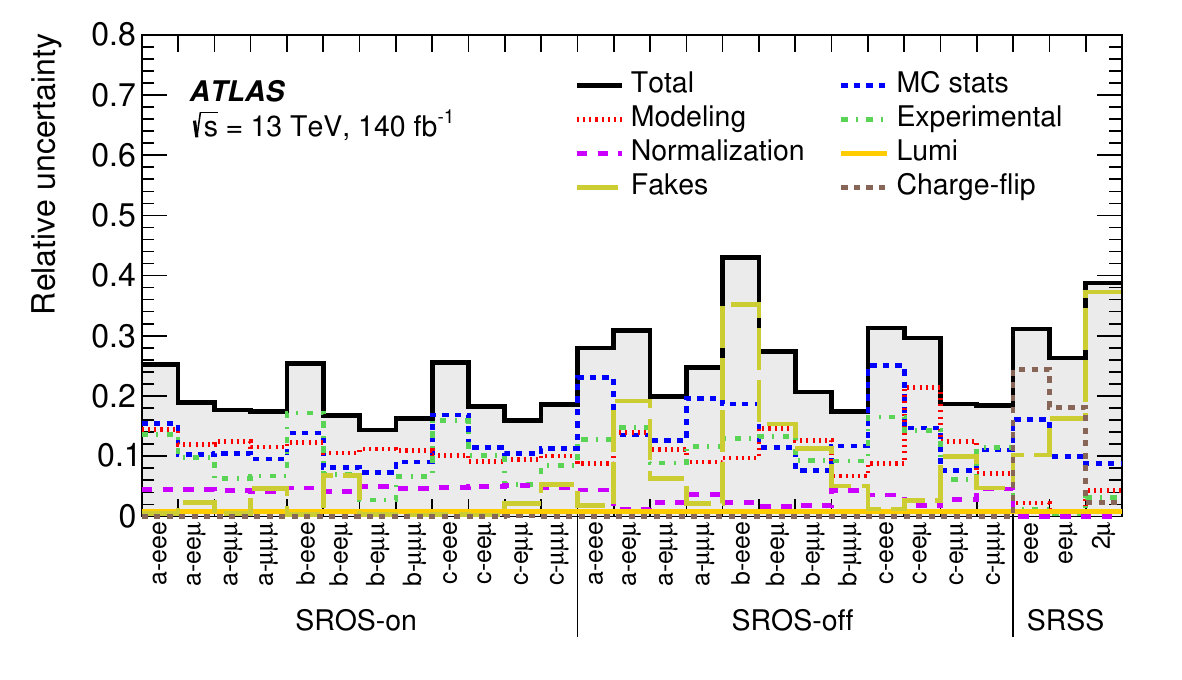}
\caption{Breakdown of the total systematic uncertainties in the background prediction for all SR channels after the background-only fit, relative to the total expected background.
The `MC stats' stands for the statistical uncertainties due to a limited number of simulated events.
The `Modeling' uncertainty groups the uncertainties due to the theoretical uncertainties, including the \WZ transfer factor uncertainties.
Uncertainties related to the experimental effects are grouped as `Experimental' uncertainty.
The `Normalization' category includes the statistical uncertainty of the data count in \CRWZ and the uncertainty from the fitted \WZ normalization factor.
The `Lumi' stands for the luminosity uncertainty.
The `Fakes' represents the uncertainties for the fake background estimation for which the bin-to-bin fluctuations are due to the small number of events in the anti-ID regions.
The uncertainties in the charge-flip background estimation are grouped as `Charge-flip' uncertainty.
Individual uncertainty categories can be (anti-)correlated.
}
\label{fig:SRsys}
\end{figure}

The MC generator modeling uncertainties for the dominant background processes, \WZ, \ZZ, and \ttbar, and the signal processes are also evaluated.
For the \WZ background, which is normalized to data in \CRWZ, these uncertainties are implemented as transfer factor uncertainties that reflect differences in the SR-to-CR or VR-to-CR ratio of yields, and therefore provide an uncertainty in the assumed shape of MC distributions across analysis regions.
The uncertainties due to the choice of the renormalization and factorization scales~\cite{Bothmann:2016nao} are assessed by varying the relevant generator parameters up and down by a factor of two with respect to the nominal values.
Uncertainties related to these variations, the choice of PDF sets~\cite{Butterworth:2015oua}, and the strong coupling constant (\alphas) value are derived by taking the envelope of the variation in the event yields~\cite{Butterworth:2015oua}.
For the \WZ and \ZZ backgrounds, uncertainties due to the virtual NLO electroweak corrections are assesed by varying the nominal scheme (\textit{additive}) to alternative schemes (\textit{multiplicative} and \textit{exponentiated}).
Additional shape uncertainties are related to the assumptions made in the event generators and PS models.
For the \WZ and \ZZ backgrounds, the \SHERPA parameters related to the PS matching scale and resummation scale are varied up and down by a factor of two with respect to the nominal values.
For the \ttbar background, the uncertainties related to the modeling of the hard scatter and to the PS are derived by comparing the \PYTHIA[8] and \HERWIG[7] predictions,
and the modeling uncertainties in the amount of initial-state radiation (ISR) and final-state radiation (FSR) are assessed by varying the related generator parameters.
The shape uncertainty in the signals are estimated by varying by a factor of two the \MGNLO parameters corresponding to the renormalization, factorization, and CKKW-L matching scales, as well as the \PYTHIA[8] shower tune parameters.

Other uncertainties include the MC statistical uncertainties, which are the largest contribution to the systematic uncertainty, mainly from the limited size of the \WZ background sample.
The uncertainties associated to the \WZ normalization factor includes the statistical uncertainty of the data count in \CRWZ.
The uncertainties related to the data-driven estimation of the charge-flip and fake backgrounds are separately evaluated, and detailed in Section~\ref{subsec:chargeflip} and \ref{subsec:fake}.
A summary of systematic uncertainties in the predicted background event yields in the SRs is shown in Figure~\ref{fig:SRsys}.

%


%

%
\section{Results}
\label{sec:result}

\subsection{Statistical analysis}
\label{subsec:StatAna}
The determination of the background and signal event yields is done through a profile log-likelihood fit~\cite{Cowan:2010js} simultaneously in the \CRWZ and all SR channels relevant to a given interpretation, using the \textsc{HistFitter}~\cite{Baak:2014wma} framework.
Systematic uncertainties are treated as Gaussian-distributed nuisance parameters in the likelihood, while the statistical uncertainties of the MC samples are treated as Poisson-distributed nuisance parameters.
The experimental uncertainties are treated as correlated between regions and processes.
The theoretical uncertainties in the background and signal predictions are treated as correlated between regions but as process independent.
The statistical uncertainties due to limited MC or data statistics are considered uncorrelated between regions and processes.

Three types of fit configurations are used to derive the results.
\begin{itemize}
\item A `\textit{background-only fit}' is performed considering only the \CRWZ and assuming no signal contribution. The normalization of the \WZ background is allowed to float and is constrained by the fit using the data in the CR. The \WZ normalization factor and nuisance parameters are profiled by maximizing the likelihood.
\item A `\textit{discovery fit}' performs the hypothesis test for a generic beyond-the-SM (BSM) signal, setting an upper limit on the number of events and visible cross-section for the signal. The fit considers only a single flavor-merged inclusive SR at once. The \CRWZ is also added to the fit to constrain the \WZ normalization if the SR belongs to \SRSFOS.
The signal contribution is allowed only in the SR, and the signal-strength parameter is bounded to be positive.
\item An `\textit{exclusion fit}' is performed to derive the exclusion limit for a given signal hypothesis. The SR channels and \CRWZ are fit simultaneously to determine the \WZ normalization factor and constrain the systematic uncertainties. The signal contamination in the \CRWZ is also taken into account according to the model predictions.
\end{itemize}

For each discovery or exclusion fit, the compatibility of the observed data with the background-only or signal-plus-background hypotheses is quantified by calculating a one-sided \(p\)-value with the profile likelihood ratio used as a test statistic~\cite{Cowan:2010js}.
The limits are derived using the \CLs prescription~\cite{Read:2002hq} where the 95\% confidence level (CL) exclusion is defined by \CLs \(< 0.05\).

\subsection{Event yields in the signal regions}
The observed data yields in each SR channel together with their SM background expectations, determined after the background-only fit, are summarized in Table~\ref{tab:SR3LSFOSonYield}, \ref{tab:SR3LSFOSoffYield}, and \ref{tab:SR3LSSYield} as well as visualised in Figure~\ref{fig:SRYield}.
No significant deviation from the SM background prediction is found in any of the SR channels.
The maximum deviation of the data from the background expectation is in \SRSFOSon-\texttt{b}-\(eee\) with a local \(1.8\sigma\) data excess, followed by a local \(1.7\sigma\) excess in \SRSFOSon-\texttt{b}-\(e\mu\mu\); the significances are computed following the prescription in Ref.~\cite{Cousins:2007bmb}.
One event is found in \SRSS-\(2\mu\), whose lepton-flavor combination is \(e\mu\mu\).

Post-fit distributions, after the background-only fit, of the key kinematic variables are shown in Figure~\ref{fig:SRdistribution}.
In particular, Figure~\ref{fig:SRdistribution}\subref{fig:SRSFOSonbemmMET} and Figure~\ref{fig:SRdistribution}\subref{fig:SRSFOSonbemmmTmin} show the \met and \mTmin distribution in \SRSFOSon-\texttt{b}-\(e\mu\mu\).
The overflow bin in \mTmin distribution, \(\mTmin\in[275,+\infty)\), which has the largest signal contribution shows a good agreement with the SM prediction and the excess in \met distribution is only seen in \(\met\in[150, 175)~\GeV\) bin, suggesting that this small excess is unlikely to be explained by the targeted SBH model.

To illustrate the sensitivity to various SBH signals throughout the regions, the MC predictions of representative simplified signal points are overlaid in the Figures.
The sensitivity to selectron signals is driven by the \(eee\) and \(e\mu\mu\) channel in \SRSFOSon, \(eee\) and \(ee\mu\) channel in \SRSFOSoff, and \(eee\) and \(ee\mu\) channel in \SRSS.
On the other hand, the sensitivity to smuon signals is dominated by the \(ee\mu\) and \(\mu\mu\mu\) channel in \SRSFOSon, \(e\mu\mu\) and \(\mu\mu\mu\) channel in \SRSFOSoff, and \(2\mu\) channel in \SRSS.
The \SRSS-\(2\mu\) shows an especially significant signal-to-background ratio for smuon signals.

\begin{table}[p!]
\centering
\caption{Observed and expected yields after the background-only fit in \SRSFOSon.
The `Others' category contains the production of Higgs boson, 3-top, 4-top, and single-top  processes.
The combined statistical and systematic uncertainties are presented.}
\resizebox{\textwidth}{!}{
\begin{tabular}{lr@{$\ \pm\ $}lr@{$\ \pm\ $}lr@{$\ \pm\ $}lr@{$\ \pm\ $}l}
\toprule
Region          & \multicolumn{2}{c}{\SRSFOSon-\texttt{a}-\(eee\)}	 & \multicolumn{2}{c}{\SRSFOSon-\texttt{a}-\(ee\mu\)}	 & \multicolumn{2}{c}{\SRSFOSon-\texttt{a}-\(e\mu\mu\)}	 & \multicolumn{2}{c}{\SRSFOSon-\texttt{a}-\(\mu\mu\mu\)}	   \\
\midrule
Observed data          & \multicolumn{2}{c}{$0$}	 & \multicolumn{2}{c}{$1$}	 & \multicolumn{2}{c}{$2$}	 & \multicolumn{2}{c}{$6$}	       \\
\midrule
Fitted SM          & $1.0$ & $0.3$          & $2.1$ & $0.4$          & $2.6$ & $0.5$          & $3.4$ & $0.6$              \\
\midrule
$WZ$	 & $0.77$ & $0.20$          & $1.6$ & $0.3$          & $2.0$ & $0.4$          & $2.4$ & $0.5$         	 \\
$ZZ$	 & \multicolumn{2}{c}{$< 0.005$}          & $0.090$ & $0.034$          & \multicolumn{2}{c}{$< 0.005$}          & $0.14$ & $0.05$         	 \\
$VVV$	 & $0.17$ & $0.13$          & $0.26$ & $0.19$          & $0.47$ & $0.28$          & $0.40$ & $0.31$         	 \\
\ttbar	 & \multicolumn{2}{c}{$< 0.05$}          & \multicolumn{2}{c}{$< 0.05$}          & $0.04$ & $_{0.04}^{0.04}$          & \multicolumn{2}{c}{$< 0.05$}         	 \\
$\ttbar X$	 & $0.059$ & $0.029$          & $0.14$ & $0.06$          & $0.13$ & $0.05$          & $0.28$ & $0.09$         	 \\
Fakes	 & \multicolumn{2}{c}{$< 0.005$}          & $0.00$ & $_{0.00}^{0.05}$          & $0.000$ & $_{0.000}^{0.014}$          & $0.11$ & $_{0.11}^{0.15}$         	 \\
Others	 & $0.019$ & $_{0.019}^{0.021}$          & \multicolumn{2}{c}{$< 0.007$}          & $0.011$ & $0.005$          & \multicolumn{2}{c}{$< 0.006$}         	 \\
\bottomrule
\toprule
Region          & \multicolumn{2}{c}{\SRSFOSon-\texttt{b}-\(eee\)}	 & \multicolumn{2}{c}{\SRSFOSon-\texttt{b}-\(ee\mu\)}	 & \multicolumn{2}{c}{\SRSFOSon-\texttt{b}-\(e\mu\mu\)}	 & \multicolumn{2}{c}{\SRSFOSon-\texttt{b}-\(\mu\mu\mu\)}	   \\
\midrule
Observed data          & \multicolumn{2}{c}{$4$}	 & \multicolumn{2}{c}{$4$}	 & \multicolumn{2}{c}{$9$}	 & \multicolumn{2}{c}{$5$}	       \\
\midrule
Fitted SM          & $1.3$ & $0.3$          & $2.7$ & $0.5$          & $4.6$ & $0.7$          & $3.9$ & $0.6$              \\
\midrule
$WZ$	 & $1.0$ & $0.3$          & $1.9$ & $0.3$          & $3.9$ & $0.6$          & $3.1$ & $0.6$         	 \\
$ZZ$	 & \multicolumn{2}{c}{$< 0.005$}          & $0.11$ & $0.04$          & $0.012$ & $0.009$          & $0.21$ & $0.06$         	 \\
$VVV$	 & $0.18$ & $0.13$          & $0.34$ & $0.23$          & $0.46$ & $0.28$          & $0.42$ & $0.30$         	 \\
\ttbar	 & \multicolumn{2}{c}{$< 0.05$}          & $0.04$ & $_{0.04}^{0.04}$          & $0.04$ & $_{0.04}^{0.05}$          & \multicolumn{2}{c}{$< 0.05$}         	 \\
$\ttbar X$	 & $0.05$ & $0.04$          & $0.18$ & $0.05$          & $0.15$ & $0.07$          & $0.13$ & $0.06$         	 \\
Fakes	 & \multicolumn{2}{c}{$< 0.003$}          & $0.14$ & $_{0.14}^{0.19}$          & $0.000$ & $_{0.000}^{0.016}$          & \multicolumn{2}{c}{$< 0.008$}         	 \\
Others	 & \multicolumn{2}{c}{$< 0.002$}          & \multicolumn{2}{c}{$< 0.005$}          & \multicolumn{2}{c}{$< 0.006$}          & \multicolumn{2}{c}{$< 0.004$}         	 \\
\bottomrule
\toprule
Region          & \multicolumn{2}{c}{\SRSFOSon-\texttt{c}-\(eee\)}	 & \multicolumn{2}{c}{\SRSFOSon-\texttt{c}-\(ee\mu\)}	 & \multicolumn{2}{c}{\SRSFOSon-\texttt{c}-\(e\mu\mu\)}	 & \multicolumn{2}{c}{\SRSFOSon-\texttt{c}-\(\mu\mu\mu\)}	   \\
\midrule
Observed data          & \multicolumn{2}{c}{$1$}	 & \multicolumn{2}{c}{$2$}	 & \multicolumn{2}{c}{$1$}	 & \multicolumn{2}{c}{$3$}	       \\
\midrule
Fitted SM          & $0.81$ & $0.21$          & $1.9$ & $0.3$          & $2.7$ & $0.4$          & $2.2$ & $0.4$              \\
\midrule
$WZ$	 & $0.66$ & $0.19$          & $1.6$ & $0.3$          & $2.4$ & $0.4$          & $1.8$ & $0.4$         	 \\
$ZZ$	 & \multicolumn{2}{c}{$< 0.005$}          & $0.031$ & $0.028$          & $0.013$ & $0.007$          & $0.073$ & $0.028$         	 \\
$VVV$	 & $0.09$ & $0.06$          & $0.19$ & $0.10$          & $0.18$ & $0.11$          & $0.20$ & $0.13$         	 \\
\ttbar	 & \multicolumn{2}{c}{$< 0.05$}          & \multicolumn{2}{c}{$< 0.05$}          & $0.06$ & $_{0.06}^{0.08}$          & \multicolumn{2}{c}{$< 0.05$}         	 \\
$\ttbar X$	 & $0.052$ & $0.026$          & $0.036$ & $0.024$          & $0.031$ & $0.022$          & $0.011$ & $_{0.011}^{0.023}$         	 \\
Fakes	 & \multicolumn{2}{c}{$< 0.002$}          & \multicolumn{2}{c}{$< 0.004$}          & $0.02$ & $_{0.02}^{0.06}$          & $0.09$ & $_{0.09}^{0.11}$         	 \\
Others	 & \multicolumn{2}{c}{$< 0.002$}          & \multicolumn{2}{c}{$< 0.002$}          & \multicolumn{2}{c}{$< 0.001$}          & \multicolumn{2}{c}{$< 0.001$}         	 \\
\bottomrule
\end{tabular}
}
\label{tab:SR3LSFOSonYield}
\end{table}

\begin{table}[p!]
\centering
\caption{Observed and expected yields after the background-only fit in \SRSFOSoff.
The `Others' category contains the production of Higgs boson, 3-top, 4-top, and single-top processes.
The combined statistical and systematic uncertainties are presented.}
\resizebox{\textwidth}{!}{
\begin{tabular}{lr@{$\ \pm\ $}lr@{$\ \pm\ $}lr@{$\ \pm\ $}lr@{$\ \pm\ $}l}
\toprule
Region          & \multicolumn{2}{c}{\SRSFOSoff-\texttt{a}-\(eee\)}	 & \multicolumn{2}{c}{\SRSFOSoff-\texttt{a}-\(ee\mu\)}	 & \multicolumn{2}{c}{\SRSFOSoff-\texttt{a}-\(e\mu\mu\)}	 & \multicolumn{2}{c}{\SRSFOSoff-\texttt{a}-\(\mu\mu\mu\)}	   \\
\midrule
Observed data          & \multicolumn{2}{c}{$0$}	 & \multicolumn{2}{c}{$3$}	 & \multicolumn{2}{c}{$2$}	 & \multicolumn{2}{c}{$1$}	       \\
\midrule
Fitted SM          & $0.39$ & $0.11$          & $1.4$ & $0.4$          & $3.8$ & $0.8$          & $1.0$ & $0.3$              \\
\midrule
$WZ$	 & $0.30$ & $0.10$          & $0.26$ & $0.12$          & $1.5$ & $0.2$          & $0.62$ & $0.21$         	 \\
$ZZ$	 & \multicolumn{2}{c}{$< 0.005$}          & $0.027$ & $0.019$          & $0.027$ & $0.016$          & $0.053$ & $0.034$         	 \\
$VVV$	 & $0.06$ & $0.04$          & $0.31$ & $0.16$          & $0.48$ & $0.20$          & $0.15$ & $0.08$         	 \\
\ttbar	 & \multicolumn{2}{c}{$< 0.05$}          & $0.41$ & $0.24$          & $1.1$ & $0.4$          & $0.13$ & $0.10$         	 \\
$\ttbar X$	 & $0.028$ & $0.017$          & $0.08$ & $0.04$          & $0.16$ & $0.06$          & $0.046$ & $0.020$         	 \\
Fakes	 & \multicolumn{2}{c}{$< 0.007$}          & $0.28$ & $0.27$          & $0.00$ & $_{0.00}^{0.24}$          & $0.000$ & $_{0.000}^{0.022}$         	 \\
Others	 & \multicolumn{2}{c}{$< 0.01$}          & $0.032$ & $0.012$          & $0.5$ & $0.5$          & $0.0101$ & $0.0034$         	 \\
\bottomrule
\toprule
Region          & \multicolumn{2}{c}{\SRSFOSoff-\texttt{b}-\(eee\)}	 & \multicolumn{2}{c}{\SRSFOSoff-\texttt{b}-\(ee\mu\)}	 & \multicolumn{2}{c}{\SRSFOSoff-\texttt{b}-\(e\mu\mu\)}	 & \multicolumn{2}{c}{\SRSFOSoff-\texttt{b}-\(\mu\mu\mu\)}	   \\
\midrule
Observed data          & \multicolumn{2}{c}{$0$}	 & \multicolumn{2}{c}{$3$}	 & \multicolumn{2}{c}{$6$}	 & \multicolumn{2}{c}{$1$}	       \\
\midrule
Fitted SM          & $0.75$ & $0.33$          & $1.6$ & $0.4$          & $3.6$ & $0.7$          & $1.5$ & $0.3$              \\
\midrule
$WZ$	 & $0.30$ & $0.10$          & $0.45$ & $0.14$          & $1.1$ & $0.2$          & $1.1$ & $0.2$         	 \\
$ZZ$	 & \multicolumn{2}{c}{$< 0.005$}          & $0.049$ & $0.029$          & $0.034$ & $0.023$          & $0.058$ & $0.029$         	 \\
$VVV$	 & $0.17$ & $0.11$          & $0.57$ & $0.28$          & $1.1$ & $0.5$          & $0.23$ & $0.11$         	 \\
\ttbar	 & \multicolumn{2}{c}{$< 0.05$}          & $0.18$ & $0.11$          & $0.61$ & $0.25$          & \multicolumn{2}{c}{$< 0.05$}         	 \\
$\ttbar X$	 & \multicolumn{2}{c}{$< 0.01$}          & $0.15$ & $0.05$          & $0.24$ & $0.07$          & $0.064$ & $0.031$         	 \\
Fakes	 & $0.20$ & $_{0.20}^{0.27}$          & $0.15$ & $_{0.15}^{0.25}$          & $0.5$ & $0.4$          & $0.02$ & $_{0.02}^{0.08}$         	 \\
Others	 & $0.07$ & $_{0.07}^{0.11}$          & $0.040$ & $0.016$          & $0.039$ & $0.011$          & \multicolumn{2}{c}{$< 0.01$}         	 \\
\bottomrule
\toprule
Region          & \multicolumn{2}{c}{\SRSFOSoff-\texttt{c}-\(eee\)}	 & \multicolumn{2}{c}{\SRSFOSoff-\texttt{c}-\(ee\mu\)}	 & \multicolumn{2}{c}{\SRSFOSoff-\texttt{c}-\(e\mu\mu\)}	 & \multicolumn{2}{c}{\SRSFOSoff-\texttt{c}-\(\mu\mu\mu\)}	   \\
\midrule
Observed data          & \multicolumn{2}{c}{$0$}	 & \multicolumn{2}{c}{$1$}	 & \multicolumn{2}{c}{$1$}	 & \multicolumn{2}{c}{$3$}	       \\
\midrule
Fitted SM          & $0.27$ & $0.08$          & $0.96$ & $0.28$          & $3.0$ & $0.6$          & $1.9$ & $0.4$              \\
\midrule
$WZ$	 & $0.16$ & $0.07$          & $0.29$ & $0.09$          & $1.4$ & $0.2$          & $1.5$ & $0.3$         	 \\
$ZZ$	 & \multicolumn{2}{c}{$< 0.005$}          & $0.020$ & $0.013$          & $0.09$ & $0.04$          & $0.04$ & $0.04$         	 \\
$VVV$	 & $0.063$ & $0.026$          & $0.52$ & $0.25$          & $0.9$ & $0.4$          & $0.31$ & $0.15$         	 \\
\ttbar	 & \multicolumn{2}{c}{$< 0.05$}          & $0.04$ & $_{0.04}^{0.05}$          & $0.09$ & $0.07$          & \multicolumn{2}{c}{$< 0.05$}         	 \\
$\ttbar X$	 & $0.039$ & $0.029$          & $0.08$ & $0.04$          & $0.12$ & $0.05$          & $0.02$ & $_{0.02}^{0.05}$         	 \\
Fakes	 & \multicolumn{2}{c}{$< 0.003$}          & $0.000$ & $_{0.000}^{0.025}$          & $0.33$ & $0.30$          & $0.05$ & $_{0.05}^{0.09}$         	 \\
Others	 & \multicolumn{2}{c}{$< 0.002$}          & \multicolumn{2}{c}{$< 0.005$}          & \multicolumn{2}{c}{$< 0.01$}          & \multicolumn{2}{c}{$< 0.004$}         	 \\
\bottomrule
\end{tabular}
}
\label{tab:SR3LSFOSoffYield}
\end{table}

\begin{table}[p!]
\centering
\caption{Observed and expected event yields after the background-only fit in \SRSS.
The combined statistical and systematic uncertainties are presented.}
\resizebox{0.70\textwidth}{!}{
\begin{tabular}{lr@{$\ \pm\ $}lr@{$\ \pm\ $}lr@{$\ \pm\ $}lr@{$\ \pm\ $}l}
\toprule
Region          & \multicolumn{2}{c}{\SRSS-\(eee\)}	 & \multicolumn{2}{c}{\SRSS-\(e\mu\mu\)}	 & \multicolumn{2}{c}{\SRSS-\(2\mu\)}	   \\
\midrule
Observed data          & \multicolumn{2}{c}{$1$}	 & \multicolumn{2}{c}{$2$}	 & \multicolumn{2}{c}{$1$}	       \\
\midrule
Fitted SM          & $0.85$ & $0.26$          & $2.0$ & $0.5$          & $0.66$ & $0.26$              \\
\midrule
Charge-flip	 & $0.74$ & $0.25$          & $1.4$ & $0.4$          & $0.015$ & $0.006$         	 \\
Fakes	 & $0.06$ & $_{0.06}^{0.09}$          & $0.57$ & $0.33$          & $0.55$ & $0.25$         	 \\
$WZ$	 & \multicolumn{2}{c}{$< 0.01$}          & $0.032$ & $0.027$          & \multicolumn{2}{c}{$< 0.005$}         	 \\
$VVV$	 & \multicolumn{2}{c}{$< 0.01$}          & $0.012$ & $_{0.012}^{0.018}$          & $0.016$ & $_{0.016}^{0.022}$         	 \\
\ttbar	 & $0.04$ & $_{0.04}^{0.04}$          & \multicolumn{2}{c}{$< 0.05$}          & $0.08$ & $0.06$         	 \\
\bottomrule
\end{tabular}
}
\label{tab:SR3LSSYield}
\end{table}
\begin{figure}[p!]
\centering
\includegraphics[width=1.0\textwidth]{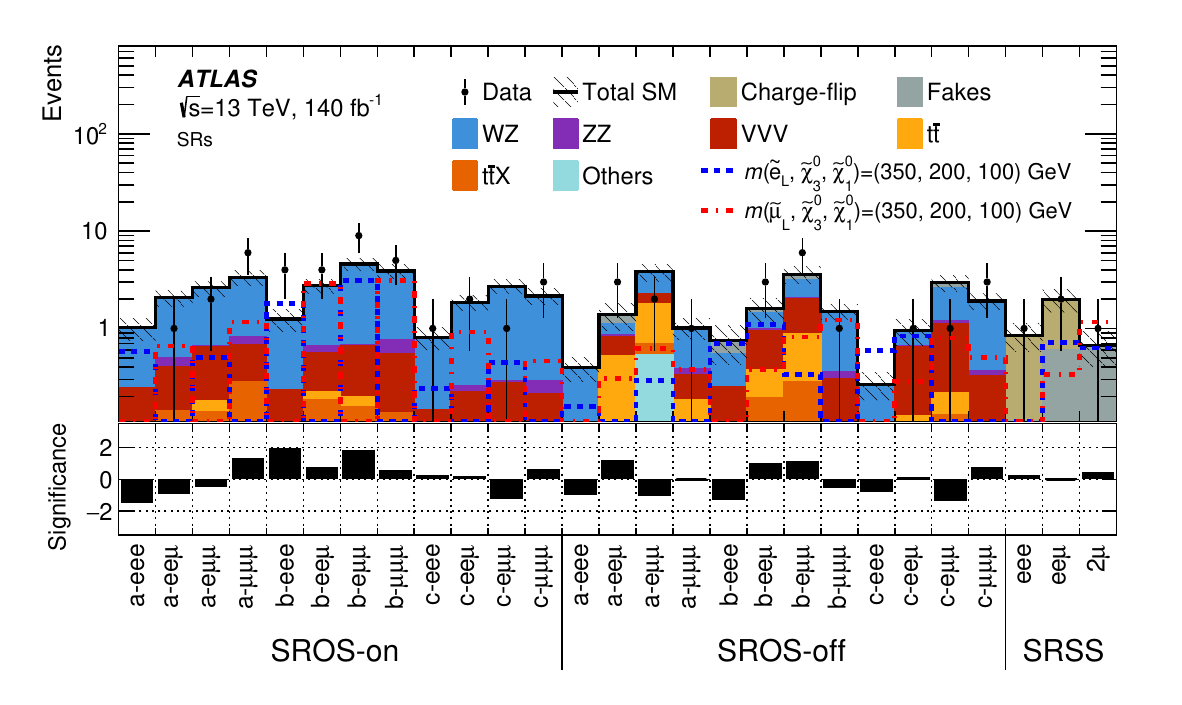}
\caption{Comparison of the observed data and the expected SM background yields in the SR channels.
The bin indices $\texttt{a},\texttt{b},\texttt{c}$ represent \(\mlll \in[30,200]\),  \(\mlll \in[200, 400)\), and  \(\mlll \in[400,+\infty)\) respectively.
The SM prediction is taken from the background-only fit.
The `Others' category contains the production of Higgs boson, 3-top, 4-top, and single-top processes.
The hatched band includes all statistical and systematic uncertainties.
Distributions for SBH \(\selL\) and \(\smuL\) signals are overlaid, with masses \(m(\sleptonL, \Nthree, \None)=(350, 200, 100)~\GeV\).
The bottom panel shows the statistical significance~\cite{Cousins:2007bmb} of the difference between the observed events and the SM expectation.
An about $1.8~\sigma$~($1.7~\sigma$) difference is obseved in \SRSFOSon-\texttt{b}-\(eee\) (\SRSFOSon-\texttt{b}-\(e\mu\mu\)).
}
\label{fig:SRYield}
\end{figure}

\begin{figure}[p!]
\centering
\subfloat[]{
\includegraphics[height=0.32\textwidth, keepaspectratio]{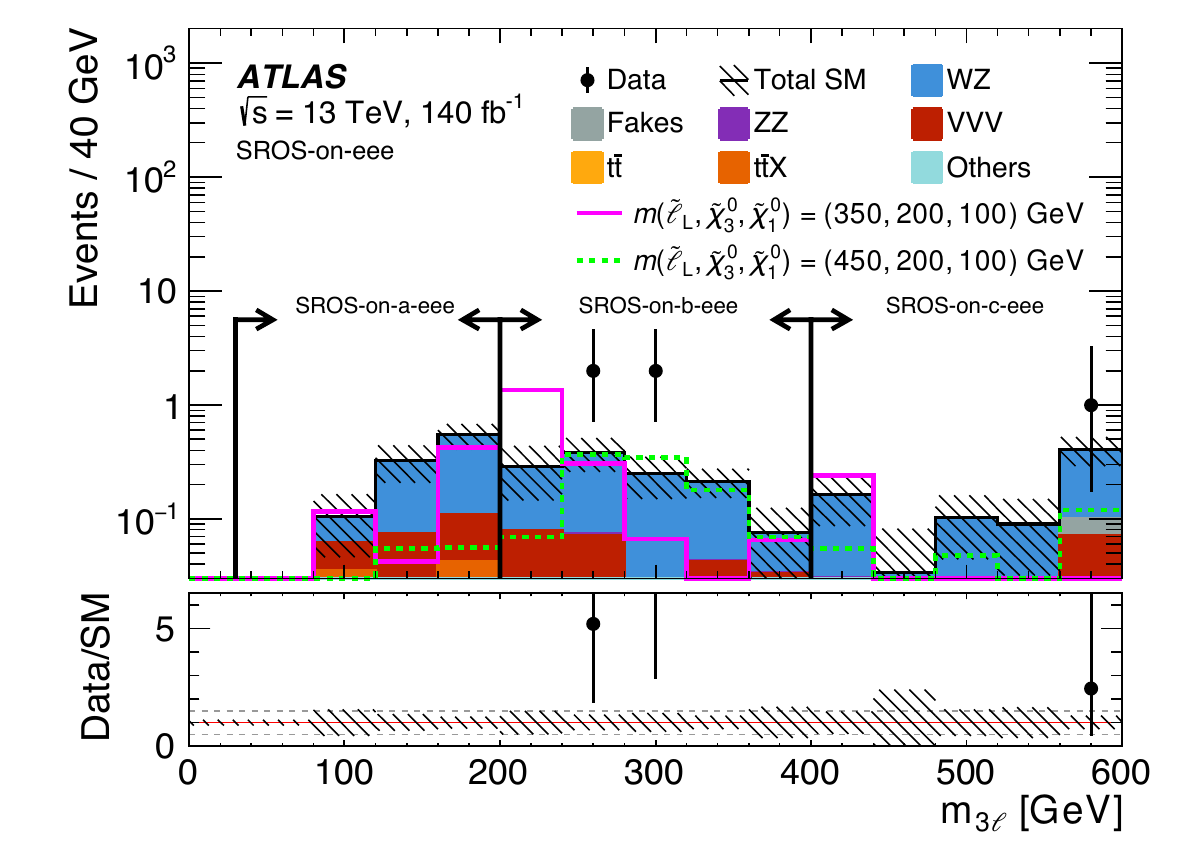}
\label{fig:SRSFOSoneeem3l}
}
\subfloat[]{
\includegraphics[height=0.32\textwidth, keepaspectratio]{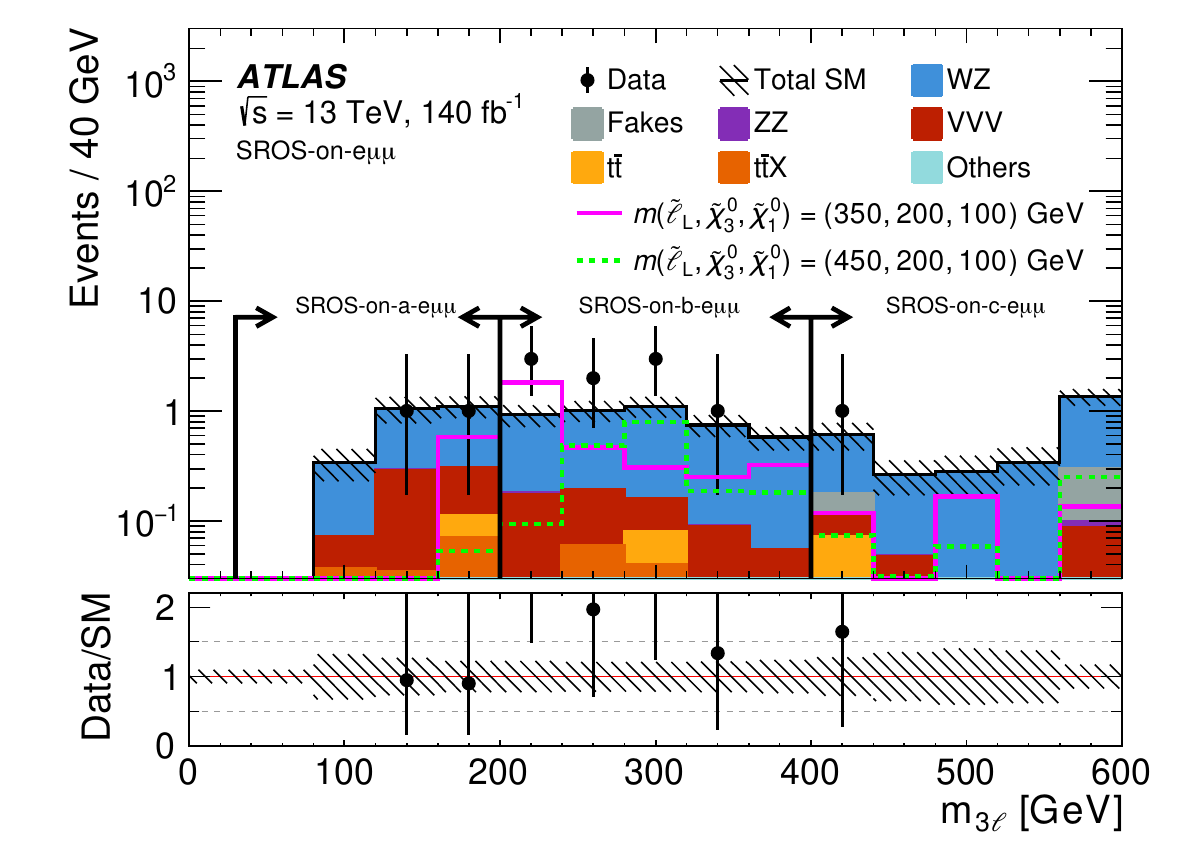}
\label{fig:SRSFOSonemmm3l}
}\\
\subfloat[]{
\includegraphics[height=0.32\textwidth, keepaspectratio]{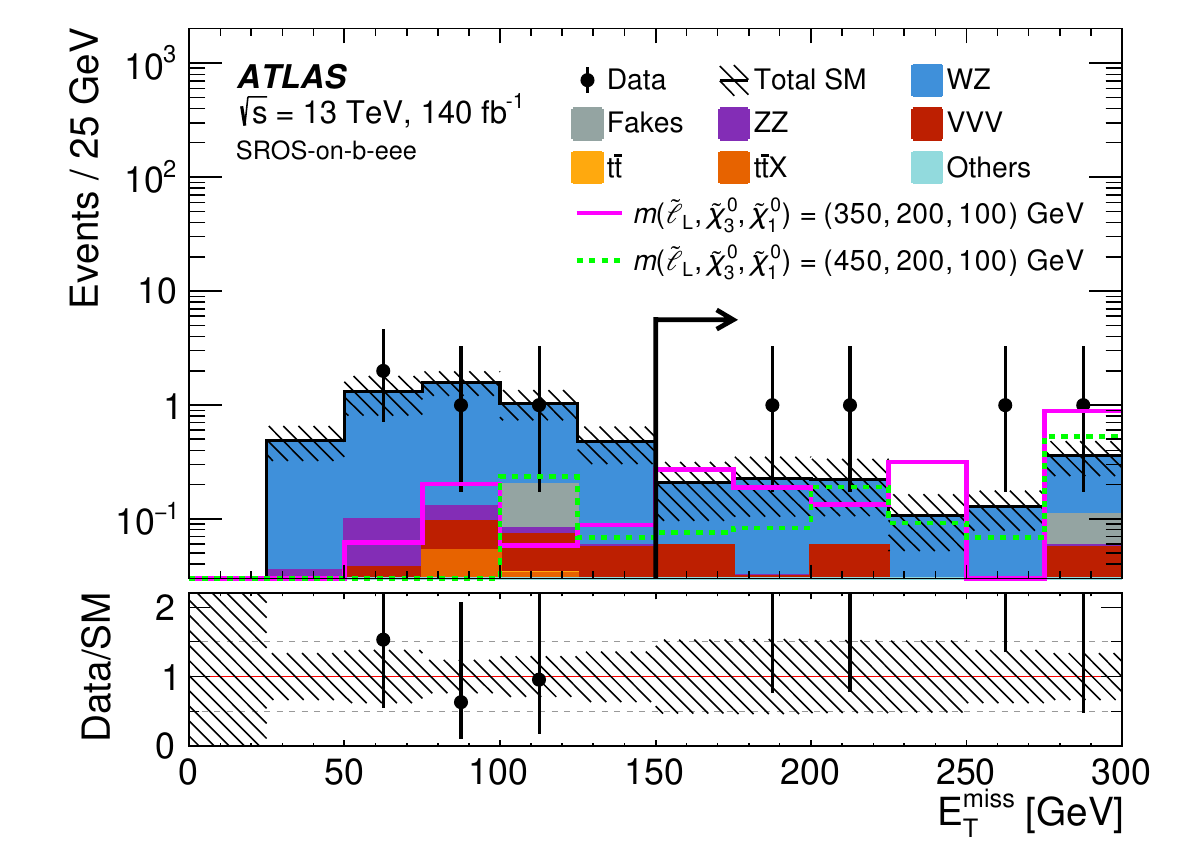}
\label{fig:SRSFOSonbeeeMET}
}
\subfloat[]{
\includegraphics[height=0.32\textwidth, keepaspectratio]{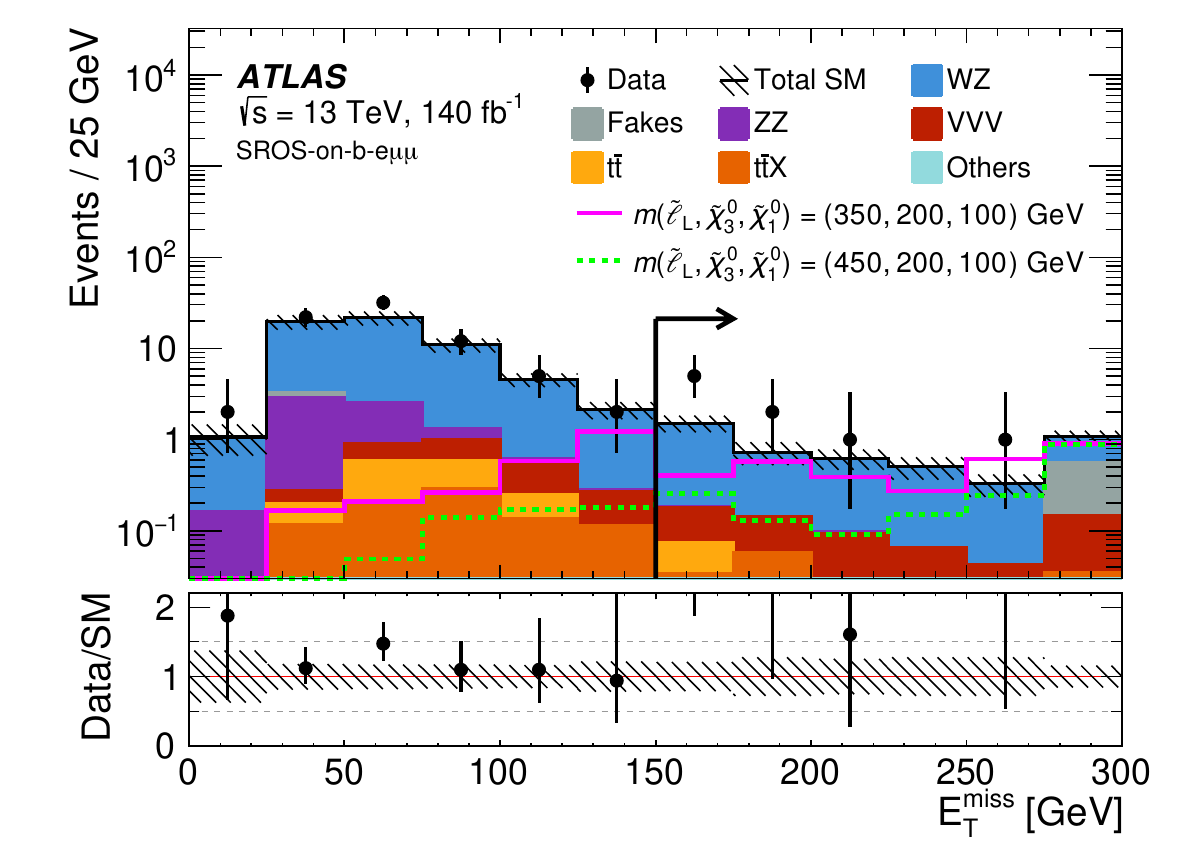}
\label{fig:SRSFOSonbemmMET}
}\\
\subfloat[]{
\includegraphics[height=0.32\textwidth, keepaspectratio]{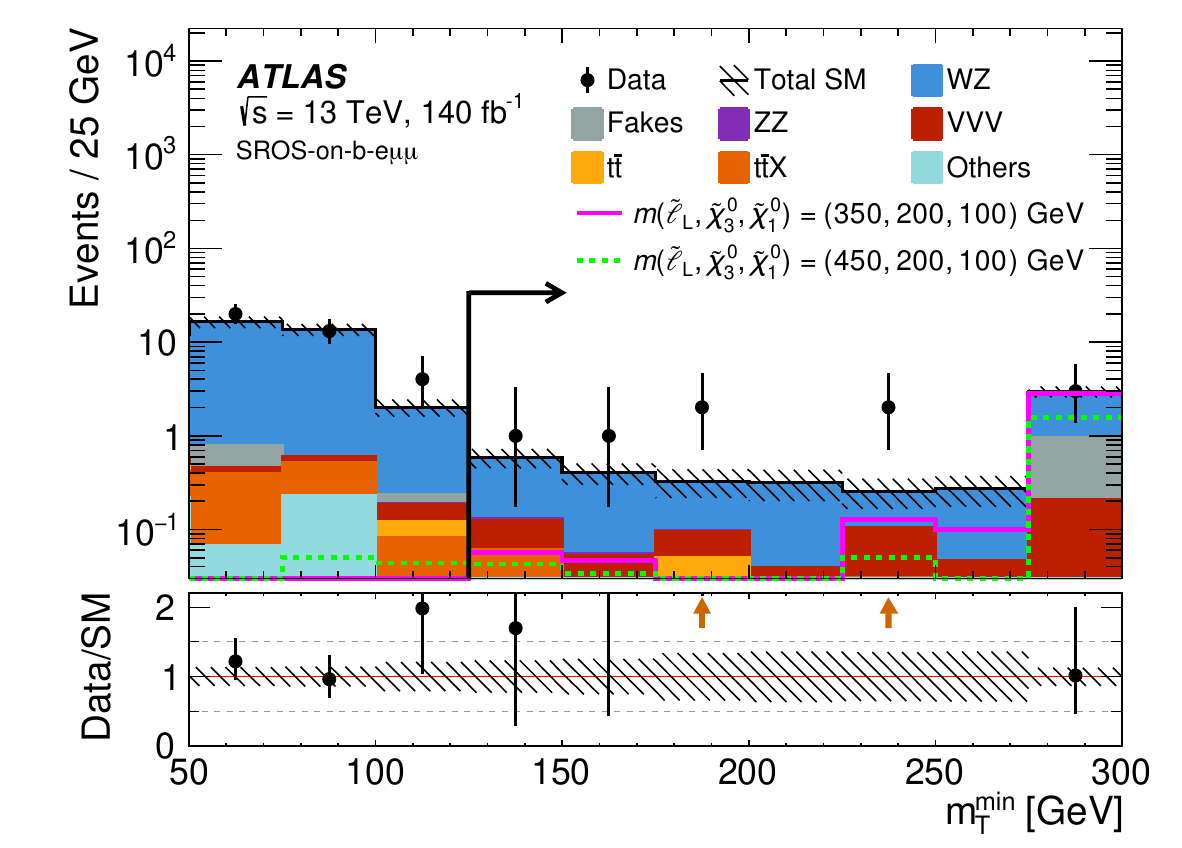}
\label{fig:SRSFOSonbemmmTmin}
}
\subfloat[]{
\includegraphics[height=0.32\textwidth, keepaspectratio]{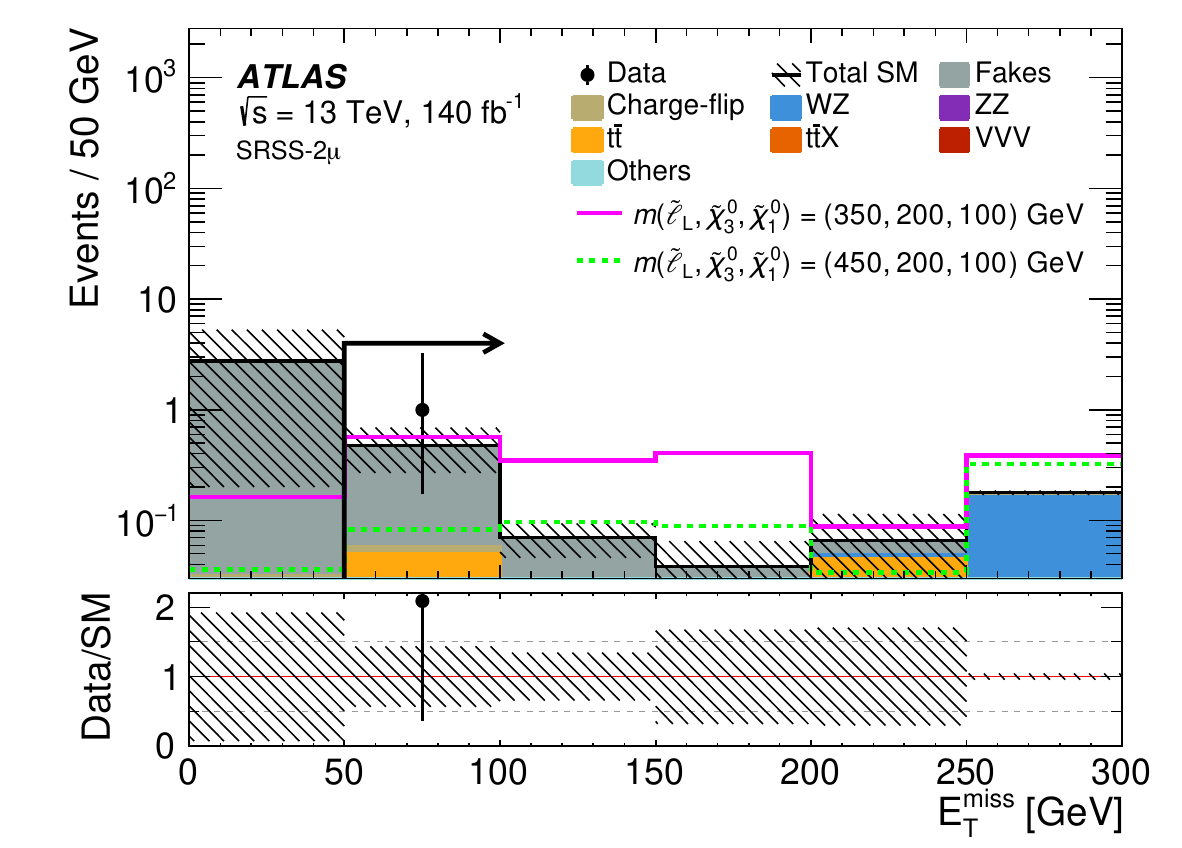}
\label{fig:SRSS2muMET}
}
\caption{Distributions of \mlll in \protect\subref{fig:SRSFOSoneeem3l} \SRSFOSon-\(eee\) and \protect\subref{fig:SRSFOSonemmm3l} \SRSFOSon-\(e\mu\mu\), of \met in \protect\subref{fig:SRSFOSonbeeeMET} \SRSFOSon-\texttt{b}-\(eee\) and \protect\subref{fig:SRSFOSonbemmMET} \SRSFOSon-\texttt{b}-\(e\mu\mu\), of \protect\subref{fig:SRSFOSonbemmmTmin} \mTmin in \SRSFOSon-\texttt{b}-\(e\mu\mu\), and of \protect\subref{fig:SRSS2muMET} \met in \SRSS-\(2\mu\).
The SR selections are applied for each distribution, except for the variable shown, for which the selection is indicated by a black arrow.
The last bin includes the overflow.
The `Others' category contains the production of Higgs boson, 3-top, 4-top, and single-top processes.
Distributions for SBH signals are overlaid.
The bottom panels show the ratio of the observed data to the predicted total background yields.
Ratio values outside the graph range are indicated by brown arrows.
The hatched band includes all statistical and systematic uncertainties.
\label{fig:SRdistribution}}
\end{figure}

\FloatBarrier

\subsection{Model-independent limits on new physics in inclusive regions}
Model-independent upper limits are derived by performing the discovery fits as described in Section~\ref{subsec:StatAna}.
The nominal flavor-binned SR channels are merged to form seven flavor-merged inclusive SRs.
The discovery fit is performed for each flavor-merged inclusive SR to derive the expected and the observed 95\% CL upper limits on the
number of the generic BSM signal events ($S_\text{exp}^{95}$ and $S_\text{obs}^{95}$) as well as the one-sided \(p\)-value of the background-only hypothesis.
Pseudo-experiments with toy MC are used for the calculation.
An upper limit on the cross-section, $\langle\epsilon{\sigma}\rangle_{\mathrm{obs}}^{95}$ where \(\epsilon\) represents the efficiency times acceptance of the flavor-merged inclusive SR for the given signal, is obtained by dividing $S_\text{obs}^{95}$ by the integrated luminosity.
The upper limits and the \(p\)-value associated with each flavor-merged inclusive SR are summarized in Table~\ref{table.results.exclxsec.pval.upperlimitToy.disc_SR.EnvelopSym}.

\begin{table}[t!]
\centering
\setlength{\tabcolsep}{0.0pc}
\caption[Breakdown of upper limits.]{
Observed ($N_{\text{obs}}$) and expected ($N_{\text{exp}}$) yields after the background-only fit for the flavor-merged inclusive SRs.
The third and fourth columns list the 95\% CL upper limits on the visible cross-section ($\sigma_{\text{vis}}^{95}$) and on the number of signal events ($S_\text{obs}^{95}$).
The fifth column ($S_\text{exp}^{95}$) shows the 95\% CL upper limit on the number of signal events, given the expected number of background events and its \(\pm 1\sigma\) variations.
The last two columns indicate the \CLb value, i.e. the confidence level observed for the background-only hypothesis, and the discovery $p$-value ($p(s = 0)$) with its associated statistical significance $Z$.
If the observed yield is below the expected yield, the $p$-value is capped at 0.5.
\label{table.results.exclxsec.pval.upperlimitToy.disc_SR.EnvelopSym}
}
\begin{tabular*}{\textwidth}{@{\extracolsep{\fill}}lccccccc}
\noalign{\smallskip}\hline\noalign{\smallskip}{Region}                       & $N_{\mathrm{obs}}$ & $N_{\mathrm{exp}}$& $\langle\epsilon{\sigma}\rangle_{\mathrm{obs}}^{95}$[fb]  &  $S_{\mathrm{obs}}^{95}$  & $S_{\mathrm{exp}}^{95}$ & \CLb & $p(s=0)$ ($Z$)  \\
\noalign{\smallskip}\hline\noalign{\smallskip}
\SRSFOSon-\texttt{a}    & $9$ & $8.94\pm1.33$ & $0.06$ &  $8.1$ & $ { 7.6 }^{ +3.1 }_{ -2.1 }$ & $0.58$ & $0.49$~$(0.01)$ \\%
\SRSFOSon-\texttt{b}    & $22$ & $12.33\pm1.67$ & $0.14$ &  $19.2$ & $ { 9.9 }^{ +3.9 }_{ -2.3 }$ & $0.98$ & $ 0.01$~$(2.24)$ \\%
\SRSFOSon-\texttt{c}    & $7$ & $7.49\pm0.99$ & $0.05$ &  $6.8$ & $ { 7.0 }^{ +2.9 }_{ -2.1 }$ & $0.46$ & $ 0.50$~$(0.00)$ \\%
\SRSFOSoff-\texttt{a}    & $6$ & $6.33\pm1.08$ & $0.04$ &  $6.1$ & $ { 6.6 }^{ +2.9 }_{ -1.9 }$ & $0.39$ & $ 0.50$~$(0.00)$ \\%
\SRSFOSoff-\texttt{b}    & $10$ & $7.47\pm1.25$ & $0.07$ &  $9.3$ & $ { 7.4 }^{ +2.8 }_{ -2.0 }$ & $0.78$ & $ 0.24$~$(0.71)$ \\%
\SRSFOSoff-\texttt{c}    & $5$ & $5.72\pm0.91$ & $0.04$ &  $5.7$ & $ { 6.2 }^{ +2.8 }_{ -1.6 }$ & $0.41$ & $ 0.50$~$(0.00)$ \\%
\SRSS    & $4$ & $3.50\pm0.78$ & $0.04$ &  $6.1$ & $ { 5.4 }^{ +2.0 }_{ -1.4 }$ & $0.65$ & $ 0.33$~$(0.43)$ \\%
\noalign{\smallskip}\hline\noalign{\smallskip}
\end{tabular*}
\end{table}

\subsection{Model-dependent exclusion limits}
The constraints on the SBH models are derived by combining all the SR channels discussed in Section~\ref{sec:selection}.
The model-dependent 95\% CL exclusion limits are calculated by performing the exclusion fits as described in Section~\ref{subsec:StatAna}.
Given the large number of signal points tested, an asymptotic approximation~\cite{Cowan:2010js} is employed in the \CLs calculation instead of the full calculation using pseudo-experiments.
The difference between the \CLs obtained from the two methods is at most 40\%, which can be translated into a difference of up to 10\% in the cross-section upper limit.

\begin{figure}[t!]
\centering
\includegraphics[width=0.65\textwidth, keepaspectratio]{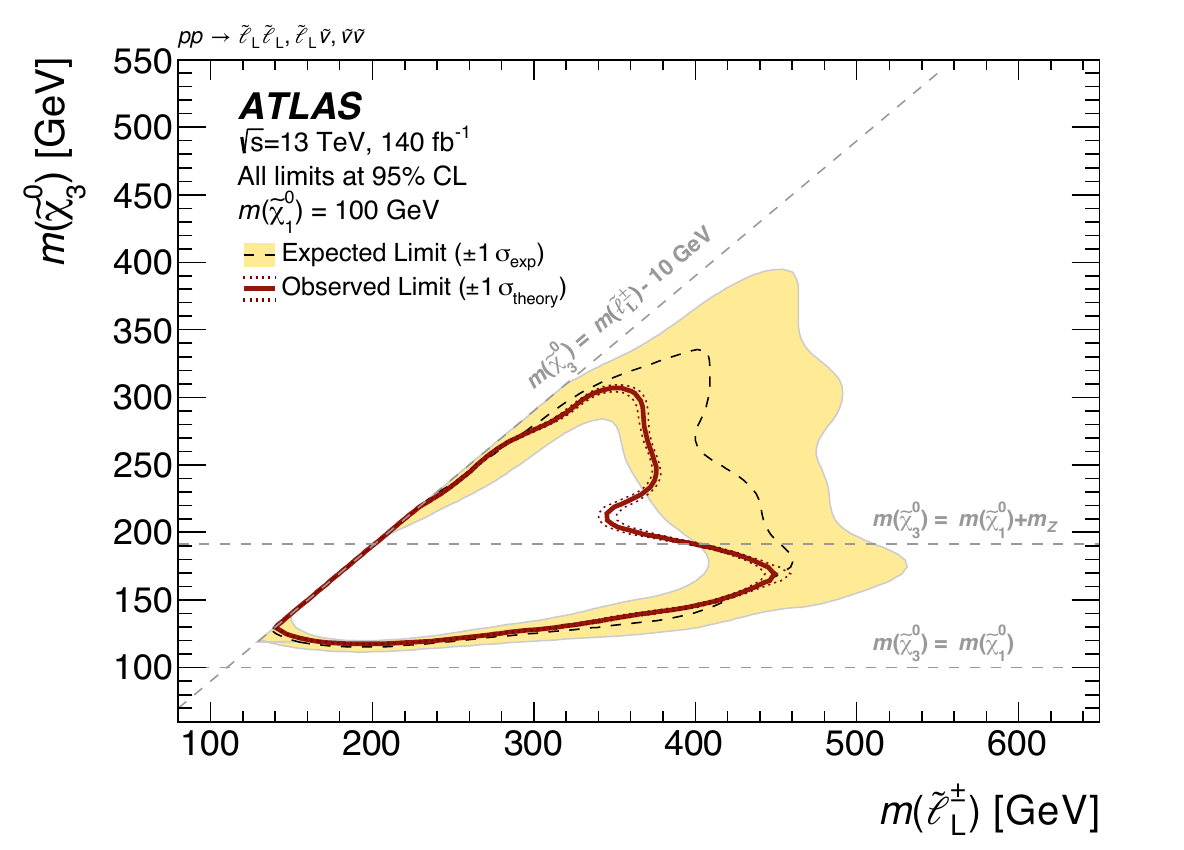}
\caption{Observed and expected exclusion limits on the SBH model where mass-degenerate \selL, \smuL and \snu are considered.
The expected 95\% CL exclusion limit is shown as a dashed black line, with the yellow band indicating \(\pm1\sigma_{\text{exp}}\) including all uncertainties except for the signal cross-section uncertainty.
The observed 95\% CL exclusion limit is shown as a red solid line, with the dotted red lines indicating \(\pm1\sigma_{\text{theory}}\) due to the signal cross-section uncertainty.
The limits are shown projected onto the \(m(\sleptonL)\) vs \(m(\Nthree)\) plane, with \(m(\None)\) assumed to be 100~\GeV.
\label{fig:ExclusionContour}
}
\end{figure}

\begin{figure}[t!]
\centering
\subfloat[]{
\includegraphics[width=0.48\textwidth, keepaspectratio]{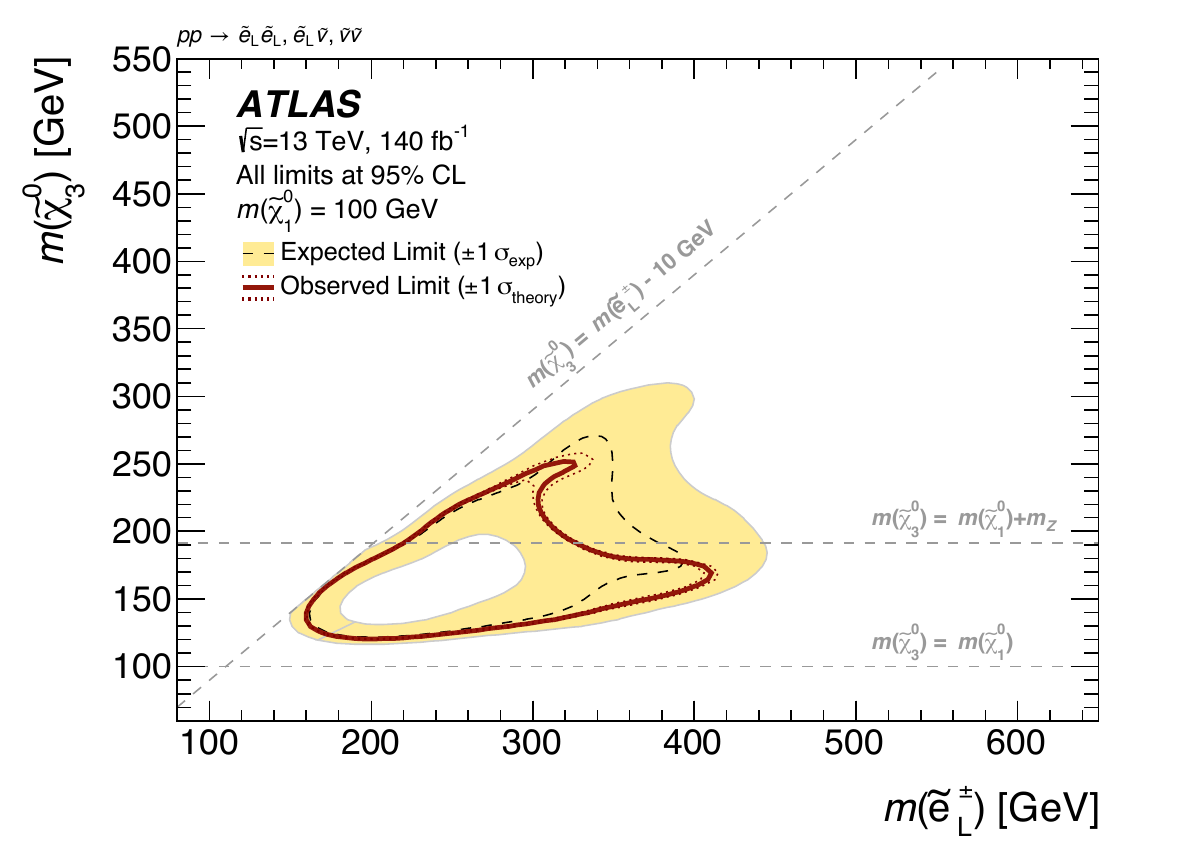} \label{fig:ExclusionContourFlavor_sele}
}
\subfloat[]{
\includegraphics[width=0.48\textwidth, keepaspectratio]{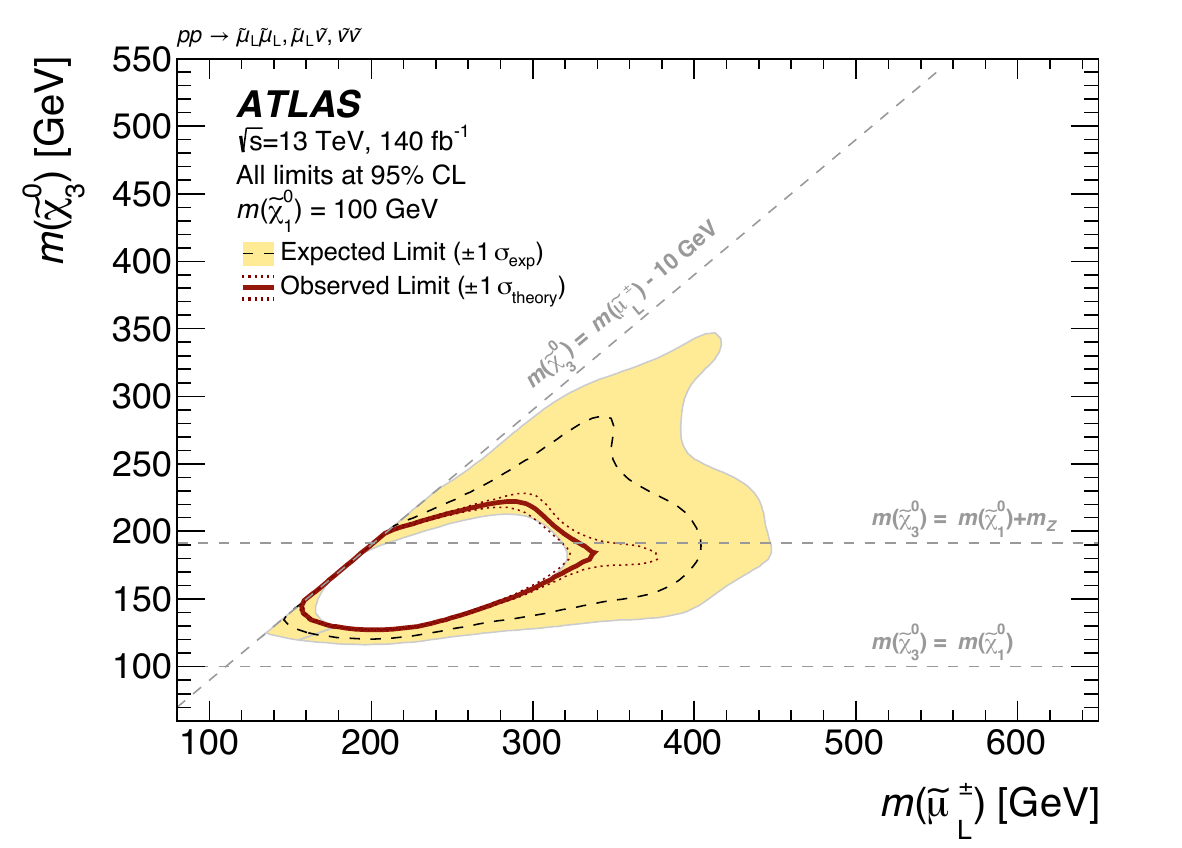} \label{fig:ExclusionContourFlavor_smu}
}
\caption{Observed and expected exclusion limits on the SBH model where only one of \protect\subref{fig:ExclusionContourFlavor_sele} \selL and \protect\subref{fig:ExclusionContourFlavor_smu} \smuL is considered and the other being decoupled.
The expected 95\% CL exclusion limit is shown as a dashed black line, with the yellow band indicating \(\pm1\sigma_{\text{exp}}\) including all uncertainties except for the signal cross-section uncertainty.
The observed 95\% CL exclusion limit is shown as a red solid line, with the dotted red lines indicating \(\pm1\sigma_{\text{theory}}\) due to the signal cross-section uncertainty.
The limits are shown projected onto the \(m(\sleptonL)\) vs \(m(\Nthree)\) plane, with \(m(\None)\) assumed to be 100~\GeV.
\label{fig:ExclusionContourFlavor}
}
\end{figure}

The \SRSFOS is mainly sensitive to models with sufficiently large \(\Delta m(\sleptonL, \Nthree)\) and \(\Delta m(\Nthree, \None)\).
On the other hand \SRSS is optimal for models with small \(\Delta m(\sleptonL, \Nthree)\), and rapidly loses sensitivity with decreasing \(\Delta m(\Nthree, \None)\).

The expected and observed exclusion limit obtained for the simplified model with mass-degenerate selectrons, smuons and sneutrinos is shown in Figure~\ref{fig:ExclusionContour}.
The observed bounds are weaker than the expected especially in \(\Delta m(\Nthree, \None) \geq m_Z\) region due to the excess of data with respect to the SM background expectation seen in \SRSFOSon-\texttt{b}-\(eee\) and \SRSFOSon-\texttt{b}-\(e\mu\mu\). %
The one-sided \(p\)-value for the background-only hypothesis with signal hypothesis \(m(\sleptonL,\Nthree,\None)=(350,200,100)~\GeV\) assuming mass-degenerate selectrons, smuons and sneutrinos is 0.07.
A left-handed slepton/sneutrino mass up to 375~\GeV is excluded in \(\Delta m(\Nthree, \None) \geq m_Z\) region and up to 450~\GeV in \(\Delta m(\Nthree, \None) < m_Z\) region at 95\% CL when \(m(\None)\) is 100~\GeV.
The exclusion limits are also set for \selL and \smuL separately by assuming that either \smuL or \selL is decoupled.
These are shown in Figure~\ref{fig:ExclusionContourFlavor}.
These results extend the sensitivity to the SBH model compared with previous searches~\cite{SUSY-2018-36,SUSY-2018-32,SUSY-2019-09}, particularly in the \(\Delta m(\Nthree,\None)\gtrsim 100~\GeV\) region which is not disfavored by direct dark matter searches.
The expected and observed exclusion limits on the \None mass are shown in Figure~\ref{fig:InterpolateCLs} onto the \(\Delta m(\sleptonL,\Nthree)\) vs \(\Delta m(\Nthree,\None)\) plane.
The observed (expected) maximum excluded \None mass is 205 (215)~\GeV for \(\Delta m(\Nthree,\None)=\Delta m(\sleptonL,\Nthree)=70~\GeV\).

\begin{figure}[t!]
\centering
\subfloat[]{
\includegraphics[width=0.48\textwidth, keepaspectratio]{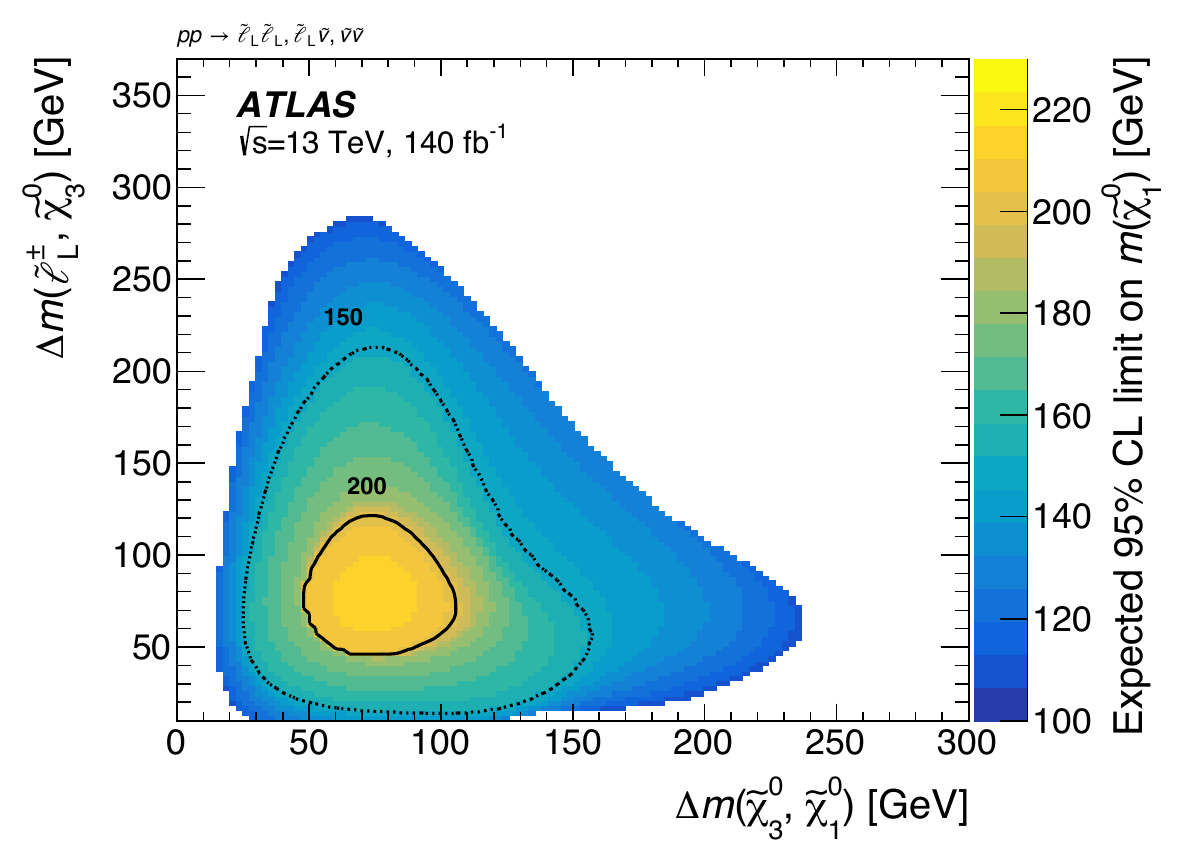} \label{fig:InterpolateCLs_exp}
}
\subfloat[]{
\includegraphics[width=0.48\textwidth, keepaspectratio]{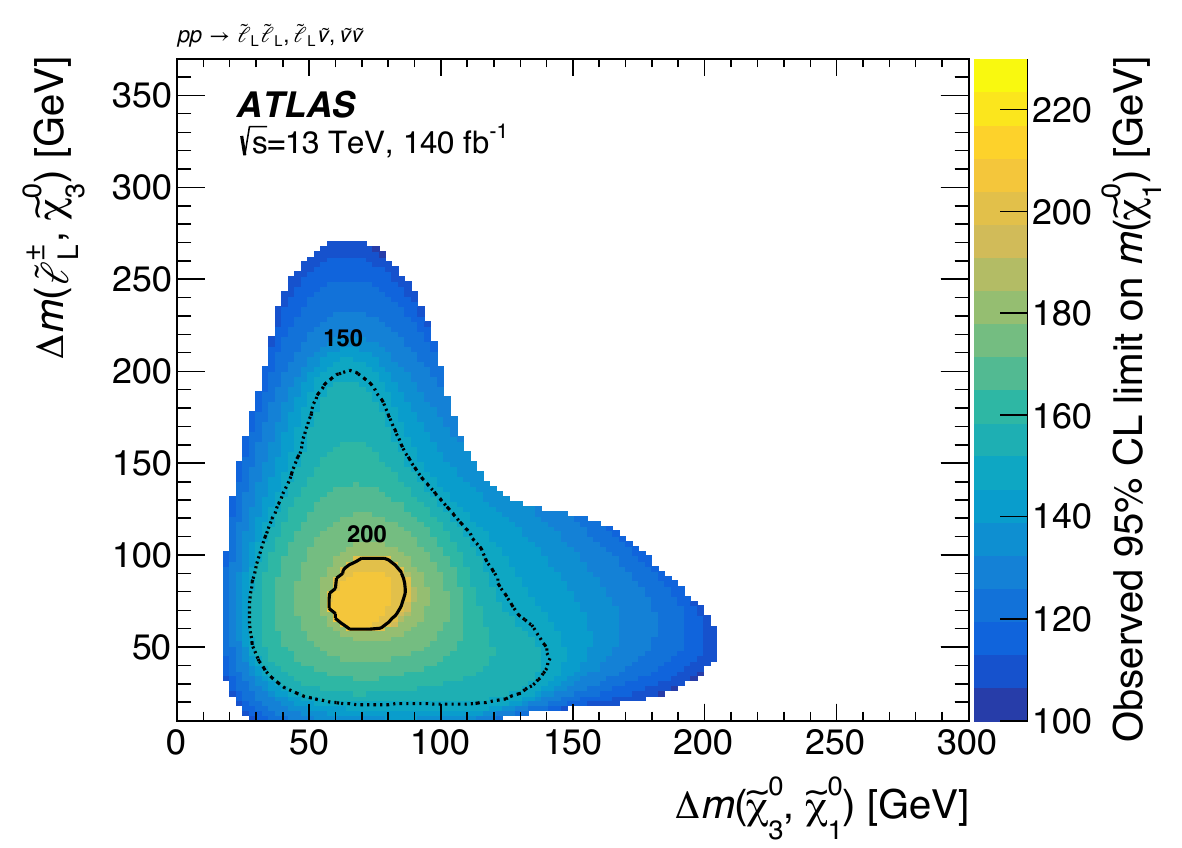} \label{fig:InterpolateCLs_obs}
}
\caption{The \protect\subref{fig:InterpolateCLs_exp} expected and \protect\subref{fig:InterpolateCLs_obs} observed lower limit on the   \None mass at 95\% CL.
The limits are shown onto the \(\Delta m(\sleptonL,\Nthree)\) vs \(\Delta m(\Nthree,\None)\) plane.
The dotted (solid) contour line indicate 150 (200)~\GeV limit of \None mass.
In the white region, the lower limit on the \None mass is below 100~\GeV.
\label{fig:InterpolateCLs}
}
\end{figure}


%

%
%
\FloatBarrier
\section{Conclusion}
\label{sec:conclusion}

This paper presents the first dedicated search for the directly produced left-handed sleptons and sneutrinos followed by a cascade decay into light higgsinos via a bino-like neutralino, motivated by the observed anomaly in the muon anomalous magnetic moment, dark matter, and electroweak naturalness arguments.
The dataset of \pp collisions at $\sqrt{s}=13~\TeV$ collected by the ATLAS experiment at the LHC from 2015 to 2018 is used, corresponding to an integrated luminosity of \(140~\ifb\).
By introducing search regions requiring three leptons with the same charge and optimizing the event selection strategy, the sensitivity to the targeted model is improved compared with previous searches, particularly in the \(\Delta m(\Nthree,\None)\gtrsim 100~\GeV\) region.
No significant deviations from the Standard Model expectations are observed and 95\% CL limits are set on the masses of relevant supersymmetric particles.
For \(m(\None)=100~\GeV\), left-handed charged slepton and sneutrino masses up to 450~\GeV are excluded assuming mass-degenerate selectrons, smuons and sneutrinos.
The highest excluded \None mass is 205~\GeV, for \(\Delta m(\Nthree,\None)=\Delta m(\sleptonL,\Nthree)=70~\GeV\).


%

%
%
%
%
%
\section*{Acknowledgments}
%

%

%
%

%
%

We thank CERN for the very successful operation of the LHC and its injectors, as well as the support staff at
CERN and at our institutions worldwide without whom ATLAS could not be operated efficiently.

The crucial computing support from all WLCG partners is acknowledged gratefully, in particular from CERN, the ATLAS Tier-1 facilities at TRIUMF/SFU (Canada), NDGF (Denmark, Norway, Sweden), CC-IN2P3 (France), KIT/GridKA (Germany), INFN-CNAF (Italy), NL-T1 (Netherlands), PIC (Spain), RAL (UK) and BNL (USA), the Tier-2 facilities worldwide and large non-WLCG resource providers. Major contributors of computing resources are listed in Ref.~\cite{ATL-SOFT-PUB-2025-001}.

We gratefully acknowledge the support of ANPCyT, Argentina; YerPhI, Armenia; ARC, Australia; BMWFW and FWF, Austria; ANAS, Azerbaijan; CNPq and FAPESP, Brazil; NSERC, NRC and CFI, Canada; CERN; ANID, Chile; CAS, MOST and NSFC, China; Minciencias, Colombia; MEYS CR, Czech Republic; DNRF and DNSRC, Denmark; IN2P3-CNRS and CEA-DRF/IRFU, France; SRNSFG, Georgia; BMBF, HGF and MPG, Germany; GSRI, Greece; RGC and Hong Kong SAR, China; ICHEP and Academy of Sciences and Humanities, Israel; INFN, Italy; MEXT and JSPS, Japan; CNRST, Morocco; NWO, Netherlands; RCN, Norway; MNiSW, Poland; FCT, Portugal; MNE/IFA, Romania; MSTDI, Serbia; MSSR, Slovakia; ARIS and MVZI, Slovenia; DSI/NRF, South Africa; MICIU/AEI, Spain; SRC and Wallenberg Foundation, Sweden; SERI, SNSF and Cantons of Bern and Geneva, Switzerland; NSTC, Taipei; TENMAK, T\"urkiye; STFC/UKRI, United Kingdom; DOE and NSF, United States of America.

Individual groups and members have received support from BCKDF, CANARIE, CRC and DRAC, Canada; CERN-CZ, FORTE and PRIMUS, Czech Republic; COST, ERC, ERDF, Horizon 2020, ICSC-NextGenerationEU and Marie Sk{\l}odowska-Curie Actions, European Union; Investissements d'Avenir Labex, Investissements d'Avenir Idex and ANR, France; DFG and AvH Foundation, Germany; Herakleitos, Thales and Aristeia programmes co-financed by EU-ESF and the Greek NSRF, Greece; BSF-NSF and MINERVA, Israel; NCN and NAWA, Poland; La Caixa Banking Foundation, CERCA Programme Generalitat de Catalunya and PROMETEO and GenT Programmes Generalitat Valenciana, Spain; G\"{o}ran Gustafssons Stiftelse, Sweden; The Royal Society and Leverhulme Trust, United Kingdom.

In addition, individual members wish to acknowledge support from Armenia: Yerevan Physics Institute (FAPERJ); CERN: European Organization for Nuclear Research (CERN DOCT); Chile: Agencia Nacional de Investigaci\'on y Desarrollo (FONDECYT 1230812, FONDECYT 1230987, FONDECYT 1240864); China: Chinese Ministry of Science and Technology (MOST-2023YFA1605700, MOST-2023YFA1609300), National Natural Science Foundation of China (NSFC - 12175119, NSFC 12275265); Czech Republic: Czech Science Foundation (GACR - 24-11373S), Ministry of Education Youth and Sports (ERC-CZ-LL2327, FORTE CZ.02.01.01/00/22\_008/0004632), PRIMUS Research Programme (PRIMUS/21/SCI/017); EU: H2020 European Research Council (ERC - 101002463); European Union: European Research Council (ERC - 948254, ERC 101089007, ERC, BARD, 101116429), European Regional Development Fund (SMASH COFUND 101081355, SLO ERDF), Horizon 2020 Framework Programme (MUCCA - CHIST-ERA-19-XAI-00), European Union, Future Artificial Intelligence Research (FAIR-NextGenerationEU PE00000013), Horizon 2020 (EuroHPC - EHPC-DEV-2024D11-051), Italian Center for High Performance Computing, Big Data and Quantum Computing (ICSC, NextGenerationEU); France: Agence Nationale de la Recherche (ANR-21-CE31-0013, ANR-21-CE31-0022, ANR-22-EDIR-0002); Germany: Baden-Württemberg Stiftung (BW Stiftung-Postdoc Eliteprogramme), Deutsche Forschungsgemeinschaft (DFG - 469666862, DFG - CR 312/5-2); China: Research Grants Council (GRF); Italy: Istituto Nazionale di Fisica Nucleare (ICSC, NextGenerationEU), Ministero dell'Università e della Ricerca (NextGenEU I53D23000820006 M4C2.1.1); Japan: Japan Society for the Promotion of Science (JSPS KAKENHI JP22H01227, JSPS KAKENHI JP22H04944, JSPS KAKENHI JP22KK0227, JSPS KAKENHI JP23KK0245); Norway: Research Council of Norway (RCN-314472); Poland: Ministry of Science and Higher Education (IDUB AGH, POB8, D4 no 9722), Polish National Science Centre (NCN 2021/42/E/ST2/00350, NCN OPUS 2023/51/B/ST2/02507, NCN OPUS nr 2022/47/B/ST2/03059, NCN UMO-2019/34/E/ST2/00393, UMO-2020/37/B/ST2/01043, UMO-2022/47/O/ST2/00148, UMO-2023/49/B/ST2/04085, UMO-2023/51/B/ST2/00920, UMO-2024/53/N/ST2/00869); Portugal: Foundation for Science and Technology (FCT); Spain: Generalitat Valenciana (Artemisa, FEDER, IDIFEDER/2018/048), Ministry of Science and Innovation (MCIN \& NextGenEU PCI2022-135018-2, MICIN \& FEDER PID2021-125273NB, RYC2019-028510-I, RYC2020-030254-I, RYC2021-031273-I, RYC2022-038164-I); Sweden: Carl Trygger Foundation (Carl Trygger Foundation CTS 22:2312), Swedish Research Council (Swedish Research Council 2023-04654, VR 2021-03651, VR 2022-03845, VR 2022-04683, VR 2023-03403, VR 2024-05451), Knut and Alice Wallenberg Foundation (KAW 2018.0458, KAW 2022.0358, KAW 2023.0366); Switzerland: Swiss National Science Foundation (SNSF - PCEFP2\_194658); United Kingdom: Leverhulme Trust (Leverhulme Trust RPG-2020-004), Royal Society (NIF-R1-231091); United States of America: U.S. Department of Energy (ECA DE-AC02-76SF00515), Neubauer Family Foundation.

%
%


%
\clearpage
\printbibliography
\clearpage
 
\begin{flushleft}
\hypersetup{urlcolor=black}
{\Large The ATLAS Collaboration}

\bigskip

\AtlasOrcid[0000-0002-6665-4934]{G.~Aad}$^\textrm{\scriptsize 104}$,
\AtlasOrcid[0000-0001-7616-1554]{E.~Aakvaag}$^\textrm{\scriptsize 17}$,
\AtlasOrcid[0000-0002-5888-2734]{B.~Abbott}$^\textrm{\scriptsize 123}$,
\AtlasOrcid[0000-0002-0287-5869]{S.~Abdelhameed}$^\textrm{\scriptsize 119a}$,
\AtlasOrcid[0000-0002-1002-1652]{K.~Abeling}$^\textrm{\scriptsize 55}$,
\AtlasOrcid[0000-0001-5763-2760]{N.J.~Abicht}$^\textrm{\scriptsize 49}$,
\AtlasOrcid[0000-0002-8496-9294]{S.H.~Abidi}$^\textrm{\scriptsize 30}$,
\AtlasOrcid[0009-0003-6578-220X]{M.~Aboelela}$^\textrm{\scriptsize 45}$,
\AtlasOrcid[0000-0002-9987-2292]{A.~Aboulhorma}$^\textrm{\scriptsize 36e}$,
\AtlasOrcid[0000-0001-5329-6640]{H.~Abramowicz}$^\textrm{\scriptsize 157}$
% ...
\bigskip
\\

% ...
$^{17}$Department for Physics and Technology, University of Bergen, Bergen; Norway.\\
% ...
$^{30}$Physics Department, Brookhaven National Laboratory, Upton NY; United States of America.\\
% ...
$^{36}$$^{(a)}$Facult\'e des Sciences Ain Chock, Universit\'e Hassan II de Casablanca;$^{(b)}$Facult\'{e} des Sciences, Universit\'{e} Ibn-Tofail, K\'{e}nitra;$^{(c)}$Facult\'e des Sciences Semlalia, Universit\'e Cadi Ayyad, LPHEA-Marrakech;$^{(d)}$LPMR, Facult\'e des Sciences, Universit\'e Mohamed Premier, Oujda;$^{(e)}$Facult\'e des sciences, Universit\'e Mohammed V, Rabat;$^{(f)}$Institute of Applied Physics, Mohammed VI Polytechnic University, Ben Guerir; Morocco.\\
% ...
$^{45}$Physics Department, Southern Methodist University, Dallas TX; United States of America.\\
% ...
$^{49}$Fakult\"{a}t Physik , Technische Universit{\"a}t Dortmund, Dortmund; Germany.\\
% ...
$^{55}$II. Physikalisches Institut, Georg-August-Universit\"{a}t G\"ottingen, G\"ottingen; Germany.\\
% ...
$^{104}$CPPM, Aix-Marseille Universit\'e, CNRS/IN2P3, Marseille; France.\\
% ...
$^{119}$$^{(a)}$New York University Abu Dhabi, Abu Dhabi;$^{(b)}$United Arab Emirates University, Al Ain; United Arab Emirates.\\
% ...
$^{123}$Homer L. Dodge Department of Physics and Astronomy, University of Oklahoma, Norman OK; United States of America.\\
% ...
$^{157}$Raymond and Beverly Sackler School of Physics and Astronomy, Tel Aviv University, Tel Aviv; Israel.\\
% ...
$^{a}$ Also at Affiliated with an institute formerly covered by a cooperation agreement with CERN.\\
$^{b}$ Also at An-Najah National University, Nablus; Palestine.\\
$^{c}$ Also at Borough of Manhattan Community College, City University of New York, New York NY; United States of America.\\
$^{d}$ Also at Center for Interdisciplinary Research and Innovation (CIRI-AUTH), Thessaloniki; Greece.\\
$^{e}$ Also at Centre of Physics of the Universities of Minho and Porto (CF-UM-UP); Portugal.\\
$^{f}$ Also at CERN, Geneva; Switzerland.\\
% ...

\end{flushleft}


%

%
%

%

%
%
%
%

\end{document}